\documentclass[aps,prb,twocolumn,superscriptaddress]{revtex4-1}

\usepackage[dvips]{graphicx}
\usepackage{color}
\usepackage[version=4]{mhchem}
\usepackage{gensymb}

\begin{document}

\title{Evolution of magnetic stripes under uniaxial stress in \ce{La_{1.885}Ba_{0.115}CuO4} studied by neutron scattering}

\author{Machteld E. Kamminga}
\email{kamminga@nbi.ku.dk}
\affiliation{Nanoscience Center, Niels Bohr Institute, University of Copenhagen, 2100 Copenhagen {\O}, Denmark}
\author{Kristine M. L. Krighaar}
\author{Astrid T. R{\o}mer}
\author{Lise {\O}. Sandberg}
\affiliation{Nanoscience Center, Niels Bohr Institute, University of Copenhagen, 2100 Copenhagen {\O}, Denmark}
\author{Pascale P. Deen}
\affiliation{Nanoscience Center, Niels Bohr Institute, University of Copenhagen, 2100 Copenhagen {\O}, Denmark}
\affiliation{European Spallation Source ERIC, Partikelgatan 224, 84 Lund, Sweden}
\author{Martin Boehm}
\affiliation{Institut Laue-Langevin, 71 Avenue des Martyrs, CS 20156, 38042 Grenoble Cedex 9,
France}
\author{G. D. Gu}
\author{J. M. Tranquada}
\affiliation{Condensed Matter Physics and Materials Science Division, Brookhaven National Laboratory,
Upton, New York 11973, USA}
\author{Kim Lefmann}
\email{lefmann@nbi.ku.dk}
\affiliation{Nanoscience Center, Niels Bohr Institute, University of Copenhagen, 2100 Copenhagen {\O}, Denmark}

\date{\today}

\begin{abstract}
Here we present the effect of uniaxial stress on the magnetic stripes in the cuprate system \ce{La_{2-x}Ba_{x}CuO4} with $x = 0.115$, previously found to have a stress-induced enhancement in the superconducting transition temperature. By means of neutron scattering, we show that the static stripes are suppressed by stress, pointing towards a trade-off between superconductivity and static magnetism, in direct agreement with previously reported $\mu$SR measurements. Additionally, we show that some of the reduced weight in the elastic channel appears to have moved to the inelastic channel. Moreover, a stress-induced momentum shift of the fluctuations towards the typical 1/8 value of commensurability is observed. These results impose a strong constraint on the theoretical interpretation of stress-enhanced superconductivity in cuprate systems.
\end{abstract}

\maketitle

\section{Introduction}
\noindent The compound \ce{La_{2-x}Ba_xCuO4} (LBCO) is famous for exhibiting pronounced spin-charge correlations known as stripes\cite{Tranquada1995,fuji04,huck11,Fradkin15}, which have been commonly observed in all hole-doped cuprate compounds.~\cite{tranquadatopo,tranquadastripe,keim15,comi16,fran20,uchi21} At a doping value of $x=1/8$, the stripes become especially pronounced and are often accompanied by a suppression in the superconducting critical temperature $T_{\rm c}$. In LBCO this suppression is particularly strong: three-dimensional superconductivity sets in at $T_{\rm c}\simeq 4$K at 1/8 doping, 
\textit{i.e.} a much lower temperature than compounds with slightly smaller or larger dopings, which display $T_{\rm c}\geq 30$ K.
However, within this ``1/8 phase'', two-dimensional superconducting fluctuations are still found up to $T_{\rm c, 2D} = 40$~K, coinciding with the temperature at which static spin stripes are observed with neutron scattering.\cite{Li2007,tranquada2008evidence}
This behaviour can be understood in terms of a spatially modulated superconducting order, the pair-density wave (PDW), which is truly two-dimensional and antagonistic to uniform $d$-wave superconductivity.~\cite{tranquadatopo,berg09b,agte20}

The tendency towards spin and charge stripe order and the intricate interplay of these types of correlations with superconductivity have motivated enormous scientific activity during the last three decades.~\cite{JULIEN2003,Fujita12,Fradkin15,tranquadastripe} It is quite clear that the two phenomena compete to a large extent, most explicitly at the 1/8 anomaly. One key question is whether there exists a way to improve superconductivity by disrupting stripe order, for example by introducing additional disorder or strain. 

In this regard, it was recently demonstrated that disorder created by proton irradiation in LBCO gave an enhanced $T_{\rm c}$.~\cite{Leroux} In a different approach, presented by Guguchia and co-authors in Ref.~\onlinecite{guguchia2020using},
a moderate uniaxial stress was applied within the CuO$_2$ crystal plane
and found to facilitate three-dimensional superconductivity at elevated temperatures in a LBCO crystal close to the 1/8 anomaly, namely for $x=0.115$.
A stress of only 25~MPa was sufficient to obtain a $T_{\rm c, onset}$ of 32~K, compared to 10~K at ambient pressure, and at 37~MPa, the stress effect is almost completely saturated. \cite{guguchia2020using}  This is in great contrast to hydrostatic pressure experiments on LBCO, which show that a pressure of a couple of GPa is needed to have a significant impact on the structure and superconducting properties.\cite{Yamada1992,Katano1993,Hucker2010} The effect of uniaxial stress on $T_{\rm c}$ in a LBCO-related stripe-ordered cuprate was first demonstrated by Takeshita \textit{et al.},\cite{Takeshita2004} showing that stress applied along the tetragonal [110] was about 3 times more efficient than along the tetragonal [100]. Notably, in the uniaxial stress study in Ref.~\onlinecite{guguchia2020using}, the magnetic volume fraction was determined by muon spin rotation ($\mu$SR) measurements and found to decrease with increasing stress, indicating that the strength of magnetic order anti-correlates with superconductivity.

In this work, we investigate the effect of uniaxial stress on LBCO $x=0.115$, similar to the crystal investigated in Ref.~\onlinecite{guguchia2020using}, by elastic and inelastic neutron scattering. We confirm the decrease of static magnetism as stress is applied along the diagonal Cu-Cu crystal axis. Furthermore, we observe that the inelastic signal at an energy transfer of 1 meV increases slightly and displays an intriguing stress-induced shift towards commensurability, \textit{i.e.} towards a periodicity of 8 lattice spacings.


\section{Experimental}

\noindent The sample was a 0.76~g single crystal of LBCO, $x=0.115$ with $T_{\rm c} \sim 12$~K, grown by the travelling-solvent floating-zone method.\cite{Tanaka89} The crystal was oriented with the [1$\overline{1}$0] direction vertical and the [0.615 0.5 0] and [001] directions in the horizontal scattering plane. The crystal was subject to uniaxial stress along the tetragonal [1$\overline{1}$0], \textit{i.e.} along the diagonal Cu-Cu direction, using a home-built pressure cell with in-situ pressure read-out, optimized for neutron scattering \cite{Sandberg2022}.

The main experiment took place on the cold neutron triple-axis spectrometer ThALES at Institut Laue-Langevin (ILL), Grenoble \cite{boehm2015thales} with the pressure cell and sample cooled by a standard Orange cryostat.  All the measurements were performed with a constant $E_{\rm f} = 5$~meV and used a double focusing monochromator and a horizontally focusing analyzer. 
All scans in reciprocal space for the elastic and inelastic magnetic stripe signals were performed as pure sample rotation scans, in order to keep the background constant.\cite{Roemer2015}

The inelastic signal was measured at the incommensurate stripe position at (0.615 0.5 2) at energy transfer 1.0~meV and 20~K, using no beam collimation. To measure the elastic signal at the (0.615 0.5 2) incommensurate position, $40'$ collimation was inserted before and after the sample to improve resolution and reduce the background at 2~K, 20~K, 45~K . The stress was applied at ambient temperatures. After that, the elastic and then the inelastic scans were repeated. Due to time constraints, however, the background measurements at 45~K were not repeated, as we did not expect any change with stress. Finally, for normalization purposes, we measured the transverse acoustic phonon around the (110) position at 2.0~meV and at 90~K. More experimental details are given in the Supplemental Information (SI).\cite{supplemental}


\section{Results and discussion}
\noindent 
After applying a uniaxial stress of 30 MPa along the tetragonal [1$\overline{1}$0] direction on our \ce{La_{1.885}Ba_{0.115}CuO4} crystal, the lattice parameters changed significantly and an orthorhombic strain was induced in the sample: the $d$-spacing of the (110) plane, measured in-plane, increased from $5.3309(8)$ \AA \ to $5.3460(3)$ \AA, a relative change of around 0.3 \%. Furthermore, the other in-plane-oriented lattice parameter, $c$, changed from $13.2764(6)$\ \AA \ to $13.1287(6)$\ \AA \ under applied stress. 



\begin{figure}[t!]
\includegraphics[width=0.48\textwidth]{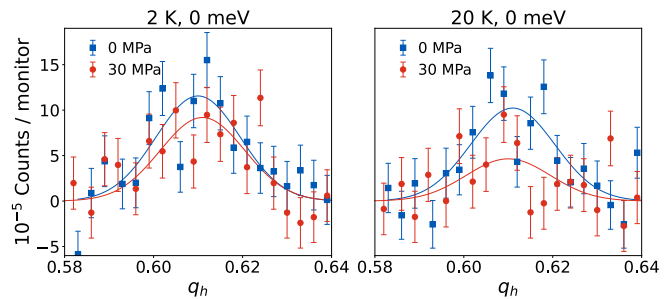}
\caption{Elastic signals at 2 K (left) and 20 K (right) without applied stress and with 30 MPa.} 
\label{Fig:elastic}
\end{figure}

In Fig.~\ref{Fig:elastic} we compare the elastic stripe peak at (0.615 0.5 2) with and without stress, measured at 2~K and 20~K. Note that 2~K is below $T_{\rm c}$, while 20~K is above at 0~MPa, but below $T_{\rm c}$ at 30~MPa.\cite{guguchia2020using} Fitting parameters and raw data without background subtraction are given in the SI.\cite{supplemental} The peak width of the elastic signals was determined by fitting the combined 2~K and 20~K data without stress (see Fig.~S1 in the SI\cite{supplemental}). As shown in Fig.~\ref{Fig:elastic}, no significant change in peak position is observed, but a clear suppression of the static magnetic signal is visible with applied stress, especially at 20~K. These results indicate a direct trade-off between spin-stripe order and 3D superconductivity. Note that this is in agreement with our expectations, since charge stripes are good for pairing, but spin-stripe fluctuations get in the way of establishing phase coherence between neighboring stripes. Moreover, spin-stripe order induces PDW superconductivity, which is primarily two dimensional due to a frustrated interlayer Josephson coupling.\cite{Tranquada1995,agte20}


\begin{figure}[b!]
\includegraphics[width=0.38\textwidth]{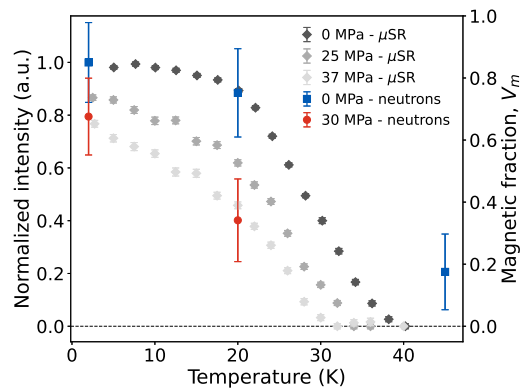}
\caption{Comparison of elastic stripe peak intensities with the magnetic volume fraction obtained by $\mu$SR, as a function of temperature and stress. The 2 K, 0 MPa, data point is normalized to a magnetic volume fraction of 85 \%, matching the 0 MPa $\mu$SR data, and the other data points are scaled accordingly. $\mu$SR data reproduced with permission from Guguchia \textit{et.\ al.} \cite{guguchia2020using}.}
\label{Fig:compare_guguchia}
\end{figure}

The suppression of the elastic spin-stripe peak by stress is in strong agreement with the $\mu$SR data presented by Guguchia \textit{et\ al.}\cite{guguchia2020using} In Fig.~\ref{Fig:compare_guguchia}, we directly compare our neutron data with the magnetic volume fractions obtained by $\mu$SR.\cite{Udby2013} Despite using different crystals, but with the same doping level, and probing the magnetism at different timescales, the results of the two independent techniques clearly show the same trend. Note that in our study, we applied the stress at an angle of 3$\degree$ from the tetragonal [1$\overline{1}$0] direction, while Guguchia \textit{et\ al.}\ applied their stress at an angle of 15$\degree$ from this direction,\cite{guguchia2020using} which would correspond to a slightly lower value of applied stress when using our geometry. Note that a small elastic intensity was found at 45~K at 0~MPa, with the error bars nearly reaching zero (see Fig.~S3 in the SI\cite{supplemental}). We argue that this residual intensity is caused by integrating over the instrumental resolution of 0.2~meV, which therefore incorporates a small fraction of the dynamic stripes that extend well beyond $T_{\rm c}$.\cite{tranquada2008evidence}

\begin{figure}[t!]
\includegraphics[width=0.34\textwidth]{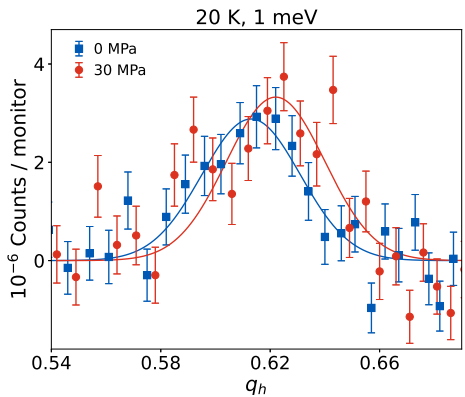}
\caption{Inelastic signals (1~meV) at 20~K without applied stress and with 30~MPa.} 
\label{Fig:inelastic}
\end{figure}

In Fig.~\ref{Fig:inelastic} we show the inelastic stripe peak at (0.615, 0.5, 2) with and without stress, measured at 20~K with an energy transfer of 1~meV. Fitting parameters and raw data without background subtraction are given in the SI.\cite{supplemental} The presence of a clear peak at 30~MPa rules out the possibility of a stress-induced spatially-uniform spin gap of 1~meV or higher. Note that LBCO outside the 1/8 anomaly, \textit{i.e.} at $x = 0.095$, also does not have a spin gap,\cite{Xu2014,Wagman2016} but our results indicate that the stress-enhanced restoration of the superconductivity to the non-1/8 level does not imply the opening of a spin gap. Furthermore, in critically doped \ce{La_{2-x}Sr_xCuO4}, \textit{i.e.}\ the Sr-analogue of LBCO, the absence of a spin gap at the 1/8 anomaly is quite robust. The elastic stripe signal is easily increased by a magnetic field, but the dynamic stripe signal is hardly affected, indicating a disconnect between the elastic and dynamic stripes in critically doped cuprates.\cite{romer2013glassy} 

In our work, the dynamic stripe intensity is not suppressed by stress, in contrast to the elastic signal. In fact, there is a hint of a slight enhancement of the peak intensity by stress. Note that large changes in intensity due to the total intensity sum rule would not be expected, since the loss of spectral weight in the elastic channel is quite small and is likely distributed over a larger energy range. However, a consistent trend is visible. These observations put a relevant constraint on the interpretations of what happens in the stress-induced $T_{\rm c}$-enhanced regime.

More strikingly, a large shift is observed in the inelastic peak position with stress: a change in $q_h$ from 0.613 $\pm$ 0.002 to 0.622 $\pm$ 0.002 is found (see Table S4 in the SI\cite{supplemental}). To make sense of this change, it may be relevant to take account of the evidence for significant spatial variations in local hole concentration provided by nuclear magnetic resonance studies on \ce{La_{2-x}Sr_xCuO4}.\cite{sing02,sing05}  We know that the average stripe wave vector varies with the average hole concentration (at least for $x < 1/8$), \cite{huck11} and it is reasonable to expect that this correlation will also occur on the local scale. Could the dynamic character depend on the local hole content and stripe wave vector?  Such a hypothesis provides a natural interpretation for the previously-observed behavior in LBCO $x = 0.095$ \cite{Xu2014} and in oxygen-doped \ce{La2CuO_{4+y}},\cite{Jacobsen2018} where the elastic spin-order peak is at an incommensurability closer to 1/8 while the low-energy spin excitations are centered at a smaller incommensurability. If regions with a local hole concentration far from 1/8 tend to have a small spin gap, while those close to 1/8 have spin-stripe order, then the former regions may be sufficient to establish three-dimensional superconducting order at a significant $T_{\rm c}$.  In the cases of LBCO $x = 0.095$ \cite{Xu2014} and \ce{La_xCuO_{4+y}},\cite{Jacobsen2018} the superconducting transitions are at 32~K and 40~K, respectively, despite the absence of a net spin gap. In the present case of LBCO $x = 0.115$, the uniaxial stress reduces the rotational anisotropy of the average structure within the \ce{CuO2} planes, thus lowering the potential for charge-stripe ordering, with the impact likely depending on the local hole density.  An inhomogeneous depression of spin-stripe order may be sufficient to yield a sharp rise in a measure of the bulk $T_{\rm c}$.\cite{guguchia2020using} Future work is required to gain insight into stress-effects on phase inhomogeneity.

\bigskip



\section{Conclusions}
\noindent In conclusion, we have shown how the magnetic stripes in \ce{La_{1.885}Ba_{0.115}CuO4} evolve under uniaxial stress along the tetragonal [1$\overline{1}$0] direction using neutron scattering. Our results show that the elastic stripes are suppressed by stress, directly matching the reduced magnetic volume fraction in the $T_{\rm c}$-enhanced regime.\cite{guguchia2020using} Furthermore, we show that stress does not open a spin gap at 1~meV and that the reduced weight in the elastic channel appears to have moved to the inelastic channel. Moreover, we observe a notable shift in the inelastic peak position towards the typical value of 1/8. Our results provide a significant constraint on the theoretical interpretation of stress-induced enhancement of $T_{\rm c}$ in LBCO, that will be of relevance to other cuprate systems as well. They support the picture of a subtle competition between spin-stripe fluctuations and superconducting phase order.


\section*{Acknowledgments}
MEK was supported by MSCA-IF Horizon 2020, grant number 838926. ATR acknowledges support from the  Independent Research Fund Denmark grant number 8021-00047. This work was supported by the Danish Natural Science Research Council through DANSCATT. The work of GG and JMT at Brookhaven is supported by the Office of Basic Energy Sciences, Materials Sciences and Engineering Division, U.S. Department of Energy under Contract No. DE-SC0012704.

 \bibliography{biblio_LBCO}

\begin{thebibliography}{34}%
\makeatletter
\providecommand \@ifxundefined [1]{%
 \@ifx{#1\undefined}
}%
\providecommand \@ifnum [1]{%
 \ifnum #1\expandafter \@firstoftwo
 \else \expandafter \@secondoftwo
 \fi
}%
\providecommand \@ifx [1]{%
 \ifx #1\expandafter \@firstoftwo
 \else \expandafter \@secondoftwo
 \fi
}%
\providecommand \natexlab [1]{#1}%
\providecommand \enquote  [1]{``#1''}%
\providecommand \bibnamefont  [1]{#1}%
\providecommand \bibfnamefont [1]{#1}%
\providecommand \citenamefont [1]{#1}%
\providecommand \href@noop [0]{\@secondoftwo}%
\providecommand \href [0]{\begingroup \@sanitize@url \@href}%
\providecommand \@href[1]{\@@startlink{#1}\@@href}%
\providecommand \@@href[1]{\endgroup#1\@@endlink}%
\providecommand \@sanitize@url [0]{\catcode `\\12\catcode `\$12\catcode
  `\&12\catcode `\#12\catcode `\^12\catcode `\_12\catcode `\%12\relax}%
\providecommand \@@startlink[1]{}%
\providecommand \@@endlink[0]{}%
\providecommand \url  [0]{\begingroup\@sanitize@url \@url }%
\providecommand \@url [1]{\endgroup\@href {#1}{\urlprefix }}%
\providecommand \urlprefix  [0]{URL }%
\providecommand \Eprint [0]{\href }%
\providecommand \doibase [0]{http://dx.doi.org/}%
\providecommand \selectlanguage [0]{\@gobble}%
\providecommand \bibinfo  [0]{\@secondoftwo}%
\providecommand \bibfield  [0]{\@secondoftwo}%
\providecommand \translation [1]{[#1]}%
\providecommand \BibitemOpen [0]{}%
\providecommand \bibitemStop [0]{}%
\providecommand \bibitemNoStop [0]{.\EOS\space}%
\providecommand \EOS [0]{\spacefactor3000\relax}%
\providecommand \BibitemShut  [1]{\csname bibitem#1\endcsname}%
\let\auto@bib@innerbib\@empty
\bibitem [{\citenamefont {Tranquada}\ \emph {et~al.}(1995)\citenamefont
  {Tranquada}, \citenamefont {Sternlieb}, \citenamefont {Axe}, \citenamefont
  {Nakamura},\ and\ \citenamefont {Uchida}}]{Tranquada1995}%
  \BibitemOpen
  \bibfield  {author} {\bibinfo {author} {\bibfnamefont {J.~M.}\ \bibnamefont
  {Tranquada}}, \bibinfo {author} {\bibfnamefont {B.~J.}\ \bibnamefont
  {Sternlieb}}, \bibinfo {author} {\bibfnamefont {J.~D.}\ \bibnamefont {Axe}},
  \bibinfo {author} {\bibfnamefont {Y.}~\bibnamefont {Nakamura}}, \ and\
  \bibinfo {author} {\bibfnamefont {S.}~\bibnamefont {Uchida}},\ }\href
  {\doibase 10.1038/375561a0} {\bibfield  {journal} {\bibinfo  {journal}
  {Nature}\ }\textbf {\bibinfo {volume} {375}},\ \bibinfo {pages} {561}
  (\bibinfo {year} {1995})}\BibitemShut {NoStop}%
\bibitem [{\citenamefont {Fujita}\ \emph {et~al.}(2004)\citenamefont {Fujita},
  \citenamefont {Goka}, \citenamefont {Yamada}, \citenamefont {Tranquada},\
  and\ \citenamefont {Regnault}}]{fuji04}%
  \BibitemOpen
  \bibfield  {author} {\bibinfo {author} {\bibfnamefont {M.}~\bibnamefont
  {Fujita}}, \bibinfo {author} {\bibfnamefont {H.}~\bibnamefont {Goka}},
  \bibinfo {author} {\bibfnamefont {K.}~\bibnamefont {Yamada}}, \bibinfo
  {author} {\bibfnamefont {J.~M.}\ \bibnamefont {Tranquada}}, \ and\ \bibinfo
  {author} {\bibfnamefont {L.~P.}\ \bibnamefont {Regnault}},\ }\href {\doibase
  10.1103/PhysRevB.70.104517} {\bibfield  {journal} {\bibinfo  {journal} {Phys.
  Rev. B}\ }\textbf {\bibinfo {volume} {70}},\ \bibinfo {pages} {104517}
  (\bibinfo {year} {2004})}\BibitemShut {NoStop}%
\bibitem [{\citenamefont {H\"ucker}\ \emph {et~al.}(2011)\citenamefont
  {H\"ucker}, \citenamefont {v.~Zimmermann}, \citenamefont {Gu}, \citenamefont
  {Xu}, \citenamefont {Wen}, \citenamefont {Xu}, \citenamefont {Kang},
  \citenamefont {Zheludev},\ and\ \citenamefont {Tranquada}}]{huck11}%
  \BibitemOpen
  \bibfield  {author} {\bibinfo {author} {\bibfnamefont {M.}~\bibnamefont
  {H\"ucker}}, \bibinfo {author} {\bibfnamefont {M.}~\bibnamefont
  {v.~Zimmermann}}, \bibinfo {author} {\bibfnamefont {G.~D.}\ \bibnamefont
  {Gu}}, \bibinfo {author} {\bibfnamefont {Z.~J.}\ \bibnamefont {Xu}}, \bibinfo
  {author} {\bibfnamefont {J.~S.}\ \bibnamefont {Wen}}, \bibinfo {author}
  {\bibfnamefont {G.}~\bibnamefont {Xu}}, \bibinfo {author} {\bibfnamefont
  {H.~J.}\ \bibnamefont {Kang}}, \bibinfo {author} {\bibfnamefont
  {A.}~\bibnamefont {Zheludev}}, \ and\ \bibinfo {author} {\bibfnamefont
  {J.~M.}\ \bibnamefont {Tranquada}},\ }\href {\doibase
  10.1103/PhysRevB.83.104506} {\bibfield  {journal} {\bibinfo  {journal} {Phys.
  Rev. B}\ }\textbf {\bibinfo {volume} {83}},\ \bibinfo {pages} {104506}
  (\bibinfo {year} {2011})}\BibitemShut {NoStop}%
\bibitem [{\citenamefont {Fradkin}\ \emph {et~al.}(2015)\citenamefont
  {Fradkin}, \citenamefont {Kivelson},\ and\ \citenamefont
  {Tranquada}}]{Fradkin15}%
  \BibitemOpen
  \bibfield  {author} {\bibinfo {author} {\bibfnamefont {E.}~\bibnamefont
  {Fradkin}}, \bibinfo {author} {\bibfnamefont {S.~A.}\ \bibnamefont
  {Kivelson}}, \ and\ \bibinfo {author} {\bibfnamefont {J.~M.}\ \bibnamefont
  {Tranquada}},\ }\href {\doibase 10.1103/RevModPhys.87.457} {\bibfield
  {journal} {\bibinfo  {journal} {Rev. Mod. Phys.}\ }\textbf {\bibinfo {volume}
  {87}},\ \bibinfo {pages} {457} (\bibinfo {year} {2015})}\BibitemShut
  {NoStop}%
\bibitem [{\citenamefont {Tranquada}(2021)}]{tranquadatopo}%
  \BibitemOpen
  \bibfield  {author} {\bibinfo {author} {\bibfnamefont {J.~M.}\ \bibnamefont
  {Tranquada}},\ }\href@noop {} {\bibfield  {journal} {\bibinfo  {journal}
  {Symmetry}\ }\textbf {\bibinfo {volume} {13}} (\bibinfo {year}
  {2021})}\BibitemShut {NoStop}%
\bibitem [{\citenamefont {Tranquada}(2020)}]{tranquadastripe}%
  \BibitemOpen
  \bibfield  {author} {\bibinfo {author} {\bibfnamefont {J.~M.}\ \bibnamefont
  {Tranquada}},\ }\href {\doibase 10.1080/00018732.2021.1935698} {\bibfield
  {journal} {\bibinfo  {journal} {Advances in Physics}\ }\textbf {\bibinfo
  {volume} {69}},\ \bibinfo {pages} {437} (\bibinfo {year} {2020})}\BibitemShut
  {NoStop}%
\bibitem [{\citenamefont {Keimer}\ \emph {et~al.}(2015)\citenamefont {Keimer},
  \citenamefont {Kivelson}, \citenamefont {Norman}, \citenamefont {Uchida},\
  and\ \citenamefont {Zaanen}}]{keim15}%
  \BibitemOpen
  \bibfield  {author} {\bibinfo {author} {\bibfnamefont {B.}~\bibnamefont
  {Keimer}}, \bibinfo {author} {\bibfnamefont {S.~A.}\ \bibnamefont
  {Kivelson}}, \bibinfo {author} {\bibfnamefont {M.~R.}\ \bibnamefont
  {Norman}}, \bibinfo {author} {\bibfnamefont {S.}~\bibnamefont {Uchida}}, \
  and\ \bibinfo {author} {\bibfnamefont {J.}~\bibnamefont {Zaanen}},\ }\href
  {\doibase 10.1038/nature14165} {\bibfield  {journal} {\bibinfo  {journal}
  {Nature}\ }\textbf {\bibinfo {volume} {518}},\ \bibinfo {pages} {179}
  (\bibinfo {year} {2015})}\BibitemShut {NoStop}%
\bibitem [{\citenamefont {Comin}\ and\ \citenamefont
  {Damascelli}(2016)}]{comi16}%
  \BibitemOpen
  \bibfield  {author} {\bibinfo {author} {\bibfnamefont {R.}~\bibnamefont
  {Comin}}\ and\ \bibinfo {author} {\bibfnamefont {A.}~\bibnamefont
  {Damascelli}},\ }\href {\doibase 10.1146/annurev-conmatphys-031115-011401}
  {\bibfield  {journal} {\bibinfo  {journal} {Annu. Rev. Condens. Matter
  Phys.}\ }\textbf {\bibinfo {volume} {7}},\ \bibinfo {pages} {369} (\bibinfo
  {year} {2016})}\BibitemShut {NoStop}%
\bibitem [{\citenamefont {Frano}\ \emph {et~al.}(2020)\citenamefont {Frano},
  \citenamefont {Blanco-Canosa}, \citenamefont {Keimer},\ and\ \citenamefont
  {Birgeneau}}]{fran20}%
  \BibitemOpen
  \bibfield  {author} {\bibinfo {author} {\bibfnamefont {A.}~\bibnamefont
  {Frano}}, \bibinfo {author} {\bibfnamefont {S.}~\bibnamefont
  {Blanco-Canosa}}, \bibinfo {author} {\bibfnamefont {B.}~\bibnamefont
  {Keimer}}, \ and\ \bibinfo {author} {\bibfnamefont {R.~J.}\ \bibnamefont
  {Birgeneau}},\ }\href {\doibase 10.1088/1361-648x/ab6140} {\bibfield
  {journal} {\bibinfo  {journal} {J. Phys. Condens. Matter}\ }\textbf {\bibinfo
  {volume} {32}},\ \bibinfo {pages} {374005} (\bibinfo {year}
  {2020})}\BibitemShut {NoStop}%
\bibitem [{\citenamefont {Uchida}(2021)}]{uchi21}%
  \BibitemOpen
  \bibfield  {author} {\bibinfo {author} {\bibfnamefont {S.-I.}\ \bibnamefont
  {Uchida}},\ }\href {\doibase 10.7566/JPSJ.90.111001} {\bibfield  {journal}
  {\bibinfo  {journal} {J. Phys. Soc. Jpn.}\ }\textbf {\bibinfo {volume}
  {90}},\ \bibinfo {pages} {111001} (\bibinfo {year} {2021})}\BibitemShut
  {NoStop}%
\bibitem [{\citenamefont {Li}\ \emph {et~al.}(2007)\citenamefont {Li},
  \citenamefont {H\"ucker}, \citenamefont {Gu}, \citenamefont {Tsvelik},\ and\
  \citenamefont {Tranquada}}]{Li2007}%
  \BibitemOpen
  \bibfield  {author} {\bibinfo {author} {\bibfnamefont {Q.}~\bibnamefont
  {Li}}, \bibinfo {author} {\bibfnamefont {M.}~\bibnamefont {H\"ucker}},
  \bibinfo {author} {\bibfnamefont {G.~D.}\ \bibnamefont {Gu}}, \bibinfo
  {author} {\bibfnamefont {A.~M.}\ \bibnamefont {Tsvelik}}, \ and\ \bibinfo
  {author} {\bibfnamefont {J.~M.}\ \bibnamefont {Tranquada}},\ }\href {\doibase
  10.1103/PhysRevLett.99.067001} {\bibfield  {journal} {\bibinfo  {journal}
  {Phys. Rev. Lett.}\ }\textbf {\bibinfo {volume} {99}},\ \bibinfo {pages}
  {067001} (\bibinfo {year} {2007})}\BibitemShut {NoStop}%
\bibitem [{\citenamefont {Tranquada}\ \emph {et~al.}(2008)\citenamefont
  {Tranquada}, \citenamefont {Gu}, \citenamefont {H{\"u}cker}, \citenamefont
  {Jie}, \citenamefont {Kang}, \citenamefont {Klingeler}, \citenamefont {Li},
  \citenamefont {Tristan}, \citenamefont {Wen}, \citenamefont {Xu},
  \citenamefont {Xu}, \citenamefont {Zhou},\ and\ \citenamefont
  {Zimmermann}}]{tranquada2008evidence}%
  \BibitemOpen
  \bibfield  {author} {\bibinfo {author} {\bibfnamefont {J.~M.}\ \bibnamefont
  {Tranquada}}, \bibinfo {author} {\bibfnamefont {G.~D.}\ \bibnamefont {Gu}},
  \bibinfo {author} {\bibfnamefont {M.}~\bibnamefont {H{\"u}cker}}, \bibinfo
  {author} {\bibfnamefont {Q.}~\bibnamefont {Jie}}, \bibinfo {author}
  {\bibfnamefont {H.-J.}\ \bibnamefont {Kang}}, \bibinfo {author}
  {\bibfnamefont {R.}~\bibnamefont {Klingeler}}, \bibinfo {author}
  {\bibfnamefont {Q.}~\bibnamefont {Li}}, \bibinfo {author} {\bibfnamefont
  {N.}~\bibnamefont {Tristan}}, \bibinfo {author} {\bibfnamefont {J.~S.}\
  \bibnamefont {Wen}}, \bibinfo {author} {\bibfnamefont {G.~Y.}\ \bibnamefont
  {Xu}}, \bibinfo {author} {\bibfnamefont {Z.~J.}\ \bibnamefont {Xu}}, \bibinfo
  {author} {\bibfnamefont {J.}~\bibnamefont {Zhou}}, \ and\ \bibinfo {author}
  {\bibfnamefont {M.~v.}\ \bibnamefont {Zimmermann}},\ }\href@noop {}
  {\bibfield  {journal} {\bibinfo  {journal} {Physical Review B}\ }\textbf
  {\bibinfo {volume} {78}},\ \bibinfo {pages} {174529} (\bibinfo {year}
  {2008})}\BibitemShut {NoStop}%
\bibitem [{\citenamefont {Berg}\ \emph {et~al.}(2009)\citenamefont {Berg},
  \citenamefont {Fradkin}, \citenamefont {Kivelson},\ and\ \citenamefont
  {Tranquada}}]{berg09b}%
  \BibitemOpen
  \bibfield  {author} {\bibinfo {author} {\bibfnamefont {E.}~\bibnamefont
  {Berg}}, \bibinfo {author} {\bibfnamefont {E.}~\bibnamefont {Fradkin}},
  \bibinfo {author} {\bibfnamefont {S.~A.}\ \bibnamefont {Kivelson}}, \ and\
  \bibinfo {author} {\bibfnamefont {J.~M.}\ \bibnamefont {Tranquada}},\
  }\href@noop {} {\bibfield  {journal} {\bibinfo  {journal} {New J. Phys.}\
  }\textbf {\bibinfo {volume} {11}},\ \bibinfo {pages} {115004} (\bibinfo
  {year} {2009})}\BibitemShut {NoStop}%
\bibitem [{\citenamefont {Agterberg}\ \emph {et~al.}(2020)\citenamefont
  {Agterberg}, \citenamefont {Davis}, \citenamefont {Edkins}, \citenamefont
  {Fradkin}, \citenamefont {Van~Harlingen}, \citenamefont {Kivelson},
  \citenamefont {Lee}, \citenamefont {Radzihovsky}, \citenamefont {Tranquada},\
  and\ \citenamefont {Wang}}]{agte20}%
  \BibitemOpen
  \bibfield  {author} {\bibinfo {author} {\bibfnamefont {D.~F.}\ \bibnamefont
  {Agterberg}}, \bibinfo {author} {\bibfnamefont {J.~S.}\ \bibnamefont
  {Davis}}, \bibinfo {author} {\bibfnamefont {S.~D.}\ \bibnamefont {Edkins}},
  \bibinfo {author} {\bibfnamefont {E.}~\bibnamefont {Fradkin}}, \bibinfo
  {author} {\bibfnamefont {D.~J.}\ \bibnamefont {Van~Harlingen}}, \bibinfo
  {author} {\bibfnamefont {S.~A.}\ \bibnamefont {Kivelson}}, \bibinfo {author}
  {\bibfnamefont {P.~A.}\ \bibnamefont {Lee}}, \bibinfo {author} {\bibfnamefont
  {L.}~\bibnamefont {Radzihovsky}}, \bibinfo {author} {\bibfnamefont {J.~M.}\
  \bibnamefont {Tranquada}}, \ and\ \bibinfo {author} {\bibfnamefont
  {Y.}~\bibnamefont {Wang}},\ }\href {\doibase
  10.1146/annurev-conmatphys-031119-050711} {\bibfield  {journal} {\bibinfo
  {journal} {Annu. Rev. Condens. Matter Phys.}\ }\textbf {\bibinfo {volume}
  {11}},\ \bibinfo {pages} {231} (\bibinfo {year} {2020})}\BibitemShut
  {NoStop}%
\bibitem [{\citenamefont {Julien}(2003)}]{JULIEN2003}%
  \BibitemOpen
  \bibfield  {author} {\bibinfo {author} {\bibfnamefont {M.-H.}\ \bibnamefont
  {Julien}},\ }\href {\doibase https://doi.org/10.1016/S0921-4526(02)01997-X}
  {\bibfield  {journal} {\bibinfo  {journal} {Physica B: Condensed Matter}\
  }\textbf {\bibinfo {volume} {329-333}},\ \bibinfo {pages} {693} (\bibinfo
  {year} {2003})}\BibitemShut {NoStop}%
\bibitem [{\citenamefont {Fujita}\ \emph {et~al.}(2012)\citenamefont {Fujita},
  \citenamefont {Hiraka}, \citenamefont {Matsuda}, \citenamefont {Matsuura},
  \citenamefont {M.~Tranquada}, \citenamefont {Wakimoto}, \citenamefont {Xu},\
  and\ \citenamefont {Yamada}}]{Fujita12}%
  \BibitemOpen
  \bibfield  {author} {\bibinfo {author} {\bibfnamefont {M.}~\bibnamefont
  {Fujita}}, \bibinfo {author} {\bibfnamefont {H.}~\bibnamefont {Hiraka}},
  \bibinfo {author} {\bibfnamefont {M.}~\bibnamefont {Matsuda}}, \bibinfo
  {author} {\bibfnamefont {M.}~\bibnamefont {Matsuura}}, \bibinfo {author}
  {\bibfnamefont {J.}~\bibnamefont {M.~Tranquada}}, \bibinfo {author}
  {\bibfnamefont {S.}~\bibnamefont {Wakimoto}}, \bibinfo {author}
  {\bibfnamefont {G.}~\bibnamefont {Xu}}, \ and\ \bibinfo {author}
  {\bibfnamefont {K.}~\bibnamefont {Yamada}},\ }\href {\doibase
  10.1143/JPSJ.81.011007} {\bibfield  {journal} {\bibinfo  {journal} {Journal
  of the Physical Society of Japan}\ }\textbf {\bibinfo {volume} {81}},\
  \bibinfo {pages} {011007} (\bibinfo {year} {2012})}\BibitemShut {NoStop}%
\bibitem [{\citenamefont {Leroux}\ \emph {et~al.}(2019)\citenamefont {Leroux},
  \citenamefont {Mishra}, \citenamefont {Ruff}, \citenamefont {Claus},
  \citenamefont {Smylie}, \citenamefont {Opagiste}, \citenamefont
  {Rodi{\`e}re}, \citenamefont {Kayani}, \citenamefont {Gu}, \citenamefont
  {Tranquada}, \citenamefont {Kwok}, \citenamefont {Islam},\ and\ \citenamefont
  {Welp}}]{Leroux}%
  \BibitemOpen
  \bibfield  {author} {\bibinfo {author} {\bibfnamefont {M.}~\bibnamefont
  {Leroux}}, \bibinfo {author} {\bibfnamefont {V.}~\bibnamefont {Mishra}},
  \bibinfo {author} {\bibfnamefont {J.~P.~C.}\ \bibnamefont {Ruff}}, \bibinfo
  {author} {\bibfnamefont {H.}~\bibnamefont {Claus}}, \bibinfo {author}
  {\bibfnamefont {M.~P.}\ \bibnamefont {Smylie}}, \bibinfo {author}
  {\bibfnamefont {C.}~\bibnamefont {Opagiste}}, \bibinfo {author}
  {\bibfnamefont {P.}~\bibnamefont {Rodi{\`e}re}}, \bibinfo {author}
  {\bibfnamefont {A.}~\bibnamefont {Kayani}}, \bibinfo {author} {\bibfnamefont
  {G.~D.}\ \bibnamefont {Gu}}, \bibinfo {author} {\bibfnamefont {J.~M.}\
  \bibnamefont {Tranquada}}, \bibinfo {author} {\bibfnamefont {W.-K.}\
  \bibnamefont {Kwok}}, \bibinfo {author} {\bibfnamefont {Z.}~\bibnamefont
  {Islam}}, \ and\ \bibinfo {author} {\bibfnamefont {U.}~\bibnamefont {Welp}},\
  }\href {\doibase 10.1073/pnas.1817134116} {\bibfield  {journal} {\bibinfo
  {journal} {Proceedings of the National Academy of Sciences}\ }\textbf
  {\bibinfo {volume} {116}},\ \bibinfo {pages} {10691} (\bibinfo {year}
  {2019})}\BibitemShut {NoStop}%
\bibitem [{\citenamefont {Guguchia}\ \emph {et~al.}(2020)\citenamefont
  {Guguchia}, \citenamefont {Das}, \citenamefont {Wang}, \citenamefont
  {Adachi}, \citenamefont {Kitajima}, \citenamefont {Elender}, \citenamefont
  {Br{\"u}ckner}, \citenamefont {Ghosh}, \citenamefont {Grinenko},
  \citenamefont {Shiroka}, \citenamefont {Müller}, \citenamefont {Mudry},
  \citenamefont {Baines}, \citenamefont {Bartkowiak}, \citenamefont {Koike},
  \citenamefont {Amato}, \citenamefont {Tranquada}, \citenamefont {Klauss},\
  and\ \citenamefont {Luetkens}}]{guguchia2020using}%
  \BibitemOpen
  \bibfield  {author} {\bibinfo {author} {\bibfnamefont {Z.}~\bibnamefont
  {Guguchia}}, \bibinfo {author} {\bibfnamefont {D.}~\bibnamefont {Das}},
  \bibinfo {author} {\bibfnamefont {C.~N.}\ \bibnamefont {Wang}}, \bibinfo
  {author} {\bibfnamefont {T.}~\bibnamefont {Adachi}}, \bibinfo {author}
  {\bibfnamefont {N.}~\bibnamefont {Kitajima}}, \bibinfo {author}
  {\bibfnamefont {M.}~\bibnamefont {Elender}}, \bibinfo {author} {\bibfnamefont
  {F.}~\bibnamefont {Br{\"u}ckner}}, \bibinfo {author} {\bibfnamefont
  {S.}~\bibnamefont {Ghosh}}, \bibinfo {author} {\bibfnamefont
  {V.}~\bibnamefont {Grinenko}}, \bibinfo {author} {\bibfnamefont
  {T.}~\bibnamefont {Shiroka}}, \bibinfo {author} {\bibfnamefont
  {M.}~\bibnamefont {Müller}}, \bibinfo {author} {\bibfnamefont
  {C.}~\bibnamefont {Mudry}}, \bibinfo {author} {\bibfnamefont
  {C.}~\bibnamefont {Baines}}, \bibinfo {author} {\bibfnamefont
  {M.}~\bibnamefont {Bartkowiak}}, \bibinfo {author} {\bibfnamefont
  {Y.}~\bibnamefont {Koike}}, \bibinfo {author} {\bibfnamefont
  {A.}~\bibnamefont {Amato}}, \bibinfo {author} {\bibfnamefont {J.~M.}\
  \bibnamefont {Tranquada}}, \bibinfo {author} {\bibfnamefont {C.~W.}\
  \bibnamefont {Klauss}, \bibfnamefont {H.-H.and~Hicks}}, \ and\ \bibinfo
  {author} {\bibfnamefont {H.}~\bibnamefont {Luetkens}},\ }\href@noop {}
  {\bibfield  {journal} {\bibinfo  {journal} {Physical Review Letters}\
  }\textbf {\bibinfo {volume} {125}},\ \bibinfo {pages} {097005} (\bibinfo
  {year} {2020})}\BibitemShut {NoStop}%
\bibitem [{\citenamefont {Yamada}\ and\ \citenamefont
  {Ido}(1992)}]{Yamada1992}%
  \BibitemOpen
  \bibfield  {author} {\bibinfo {author} {\bibfnamefont {N.}~\bibnamefont
  {Yamada}}\ and\ \bibinfo {author} {\bibfnamefont {M.}~\bibnamefont {Ido}},\
  }\href {\doibase https://doi.org/10.1016/0921-4534(92)90029-C} {\bibfield
  {journal} {\bibinfo  {journal} {Physica C: Superconductivity}\ }\textbf
  {\bibinfo {volume} {203}},\ \bibinfo {pages} {240} (\bibinfo {year}
  {1992})}\BibitemShut {NoStop}%
\bibitem [{\citenamefont {Katano}\ \emph {et~al.}(1993)\citenamefont {Katano},
  \citenamefont {Funahashi}, \citenamefont {M\^ori}, \citenamefont {Ueda},\
  and\ \citenamefont {Fernandez-Baca}}]{Katano1993}%
  \BibitemOpen
  \bibfield  {author} {\bibinfo {author} {\bibfnamefont {S.}~\bibnamefont
  {Katano}}, \bibinfo {author} {\bibfnamefont {S.}~\bibnamefont {Funahashi}},
  \bibinfo {author} {\bibfnamefont {N.}~\bibnamefont {M\^ori}}, \bibinfo
  {author} {\bibfnamefont {Y.}~\bibnamefont {Ueda}}, \ and\ \bibinfo {author}
  {\bibfnamefont {J.~A.}\ \bibnamefont {Fernandez-Baca}},\ }\href {\doibase
  10.1103/PhysRevB.48.6569} {\bibfield  {journal} {\bibinfo  {journal} {Phys.
  Rev. B}\ }\textbf {\bibinfo {volume} {48}},\ \bibinfo {pages} {6569}
  (\bibinfo {year} {1993})}\BibitemShut {NoStop}%
\bibitem [{\citenamefont {H\"ucker}\ \emph {et~al.}(2010)\citenamefont
  {H\"ucker}, \citenamefont {v.~Zimmermann}, \citenamefont {Debessai},
  \citenamefont {Schilling}, \citenamefont {Tranquada},\ and\ \citenamefont
  {Gu}}]{Hucker2010}%
  \BibitemOpen
  \bibfield  {author} {\bibinfo {author} {\bibfnamefont {M.}~\bibnamefont
  {H\"ucker}}, \bibinfo {author} {\bibfnamefont {M.}~\bibnamefont
  {v.~Zimmermann}}, \bibinfo {author} {\bibfnamefont {M.}~\bibnamefont
  {Debessai}}, \bibinfo {author} {\bibfnamefont {J.~S.}\ \bibnamefont
  {Schilling}}, \bibinfo {author} {\bibfnamefont {J.~M.}\ \bibnamefont
  {Tranquada}}, \ and\ \bibinfo {author} {\bibfnamefont {G.~D.}\ \bibnamefont
  {Gu}},\ }\href {\doibase 10.1103/PhysRevLett.104.057004} {\bibfield
  {journal} {\bibinfo  {journal} {Phys. Rev. Lett.}\ }\textbf {\bibinfo
  {volume} {104}},\ \bibinfo {pages} {057004} (\bibinfo {year}
  {2010})}\BibitemShut {NoStop}%
\bibitem [{\citenamefont {Takeshita}\ \emph {et~al.}(2004)\citenamefont
  {Takeshita}, \citenamefont {Sasagawa}, \citenamefont {Sugioka}, \citenamefont
  {Tokura},\ and\ \citenamefont {Takagi}}]{Takeshita2004}%
  \BibitemOpen
  \bibfield  {author} {\bibinfo {author} {\bibfnamefont {N.}~\bibnamefont
  {Takeshita}}, \bibinfo {author} {\bibfnamefont {T.}~\bibnamefont {Sasagawa}},
  \bibinfo {author} {\bibfnamefont {T.}~\bibnamefont {Sugioka}}, \bibinfo
  {author} {\bibfnamefont {Y.}~\bibnamefont {Tokura}}, \ and\ \bibinfo {author}
  {\bibfnamefont {H.}~\bibnamefont {Takagi}},\ }\href {\doibase
  10.1143/JPSJ.73.1123} {\bibfield  {journal} {\bibinfo  {journal} {Journal of
  the Physical Society of Japan}\ }\textbf {\bibinfo {volume} {73}},\ \bibinfo
  {pages} {1123} (\bibinfo {year} {2004})}\BibitemShut {NoStop}%
\bibitem [{\citenamefont {Tanaka}\ \emph {et~al.}(1989)\citenamefont {Tanaka},
  \citenamefont {Yamane},\ and\ \citenamefont {Kojima}}]{Tanaka89}%
  \BibitemOpen
  \bibfield  {author} {\bibinfo {author} {\bibfnamefont {I.}~\bibnamefont
  {Tanaka}}, \bibinfo {author} {\bibfnamefont {K.}~\bibnamefont {Yamane}}, \
  and\ \bibinfo {author} {\bibfnamefont {H.}~\bibnamefont {Kojima}},\ }\href
  {\doibase https://doi.org/10.1016/0022-0248(89)90074-2} {\bibfield  {journal}
  {\bibinfo  {journal} {Journal of Crystal Growth}\ }\textbf {\bibinfo {volume}
  {96}},\ \bibinfo {pages} {711} (\bibinfo {year} {1989})}\BibitemShut
  {NoStop}%
\bibitem [{\citenamefont {Sandberg~{\em et al.}}(2022)}]{Sandberg2022}%
  \BibitemOpen
  \bibfield  {author} {\bibinfo {author} {\bibfnamefont {L.~{\O}.}\
  \bibnamefont {Sandberg~{\em et al.}}},\ }\href@noop {} {\bibfield  {journal}
  {\bibinfo  {journal} {manuscript in preparation}\ } (\bibinfo {year}
  {2022})}\BibitemShut {NoStop}%
\bibitem [{\citenamefont {Boehm}\ \emph {et~al.}(2015)\citenamefont {Boehm},
  \citenamefont {Steffens}, \citenamefont {Kulda}, \citenamefont {Klicpera},
  \citenamefont {Roux}, \citenamefont {Courtois}, \citenamefont {Svoboda},
  \citenamefont {Saroun},\ and\ \citenamefont {Sechovsky}}]{boehm2015thales}%
  \BibitemOpen
  \bibfield  {author} {\bibinfo {author} {\bibfnamefont {M.}~\bibnamefont
  {Boehm}}, \bibinfo {author} {\bibfnamefont {P.}~\bibnamefont {Steffens}},
  \bibinfo {author} {\bibfnamefont {J.}~\bibnamefont {Kulda}}, \bibinfo
  {author} {\bibfnamefont {M.}~\bibnamefont {Klicpera}}, \bibinfo {author}
  {\bibfnamefont {S.}~\bibnamefont {Roux}}, \bibinfo {author} {\bibfnamefont
  {P.}~\bibnamefont {Courtois}}, \bibinfo {author} {\bibfnamefont
  {P.}~\bibnamefont {Svoboda}}, \bibinfo {author} {\bibfnamefont
  {J.}~\bibnamefont {Saroun}}, \ and\ \bibinfo {author} {\bibfnamefont
  {V.}~\bibnamefont {Sechovsky}},\ }\href@noop {} {\bibfield  {journal}
  {\bibinfo  {journal} {Neutron News}\ }\textbf {\bibinfo {volume} {26}},\
  \bibinfo {pages} {18} (\bibinfo {year} {2015})}\BibitemShut {NoStop}%
\bibitem [{\citenamefont {R{\o}mer}\ \emph {et~al.}(2015)\citenamefont
  {R{\o}mer}, \citenamefont {Ray}, \citenamefont {Jacobsen}, \citenamefont
  {Udby}, \citenamefont {Andersen}, \citenamefont {Bertelsen}, \citenamefont
  {Holm}, \citenamefont {Christensen}, \citenamefont {Toft-Petersen},
  \citenamefont {Skoulatos}, \citenamefont {Laver}, \citenamefont
  {Schneidewind}, \citenamefont {Link}, \citenamefont {Oda}, \citenamefont
  {Ido}, \citenamefont {Momono},\ and\ \citenamefont {Lefmann}}]{Roemer2015}%
  \BibitemOpen
  \bibfield  {author} {\bibinfo {author} {\bibfnamefont {A.~T.}\ \bibnamefont
  {R{\o}mer}}, \bibinfo {author} {\bibfnamefont {P.~J.}\ \bibnamefont {Ray}},
  \bibinfo {author} {\bibfnamefont {H.}~\bibnamefont {Jacobsen}}, \bibinfo
  {author} {\bibfnamefont {L.}~\bibnamefont {Udby}}, \bibinfo {author}
  {\bibfnamefont {B.~M.}\ \bibnamefont {Andersen}}, \bibinfo {author}
  {\bibfnamefont {M.}~\bibnamefont {Bertelsen}}, \bibinfo {author}
  {\bibfnamefont {S.~L.}\ \bibnamefont {Holm}}, \bibinfo {author}
  {\bibfnamefont {N.~B.}\ \bibnamefont {Christensen}}, \bibinfo {author}
  {\bibfnamefont {R.}~\bibnamefont {Toft-Petersen}}, \bibinfo {author}
  {\bibfnamefont {M.}~\bibnamefont {Skoulatos}}, \bibinfo {author}
  {\bibfnamefont {M.}~\bibnamefont {Laver}}, \bibinfo {author} {\bibfnamefont
  {A.}~\bibnamefont {Schneidewind}}, \bibinfo {author} {\bibfnamefont
  {P.}~\bibnamefont {Link}}, \bibinfo {author} {\bibfnamefont {M.}~\bibnamefont
  {Oda}}, \bibinfo {author} {\bibfnamefont {M.}~\bibnamefont {Ido}}, \bibinfo
  {author} {\bibfnamefont {N.}~\bibnamefont {Momono}}, \ and\ \bibinfo {author}
  {\bibfnamefont {K.}~\bibnamefont {Lefmann}},\ }\href@noop {} {\bibfield
  {journal} {\bibinfo  {journal} {Physical Review B}\ }\textbf {\bibinfo
  {volume} {91}},\ \bibinfo {pages} {174507} (\bibinfo {year}
  {2015})}\BibitemShut {NoStop}%
\bibitem [{sup()}]{supplemental}%
  \BibitemOpen
  \href@noop {} {}\bibinfo {note} {See Supplemental Material at [URL will be
  inserted by publisher] for experimental details, the scattering geometry, raw
  data and fitting parameters.}\BibitemShut {Stop}%
\bibitem [{\citenamefont {Udby}\ \emph {et~al.}(2013)\citenamefont {Udby},
  \citenamefont {Larsen}, \citenamefont {Christensen}, \citenamefont {Boehm},
  \citenamefont {Niedermayer}, \citenamefont {Mohottala}, \citenamefont
  {Jensen}, \citenamefont {Toft-Petersen}, \citenamefont {Chou}, \citenamefont
  {Andersen}, \citenamefont {Lefmann},\ and\ \citenamefont {Wells}}]{Udby2013}%
  \BibitemOpen
  \bibfield  {author} {\bibinfo {author} {\bibfnamefont {L.}~\bibnamefont
  {Udby}}, \bibinfo {author} {\bibfnamefont {J.}~\bibnamefont {Larsen}},
  \bibinfo {author} {\bibfnamefont {N.~B.}\ \bibnamefont {Christensen}},
  \bibinfo {author} {\bibfnamefont {M.}~\bibnamefont {Boehm}}, \bibinfo
  {author} {\bibfnamefont {C.}~\bibnamefont {Niedermayer}}, \bibinfo {author}
  {\bibfnamefont {H.~E.}\ \bibnamefont {Mohottala}}, \bibinfo {author}
  {\bibfnamefont {T.~B.~S.}\ \bibnamefont {Jensen}}, \bibinfo {author}
  {\bibfnamefont {R.}~\bibnamefont {Toft-Petersen}}, \bibinfo {author}
  {\bibfnamefont {F.~C.}\ \bibnamefont {Chou}}, \bibinfo {author}
  {\bibfnamefont {N.~H.}\ \bibnamefont {Andersen}}, \bibinfo {author}
  {\bibfnamefont {K.}~\bibnamefont {Lefmann}}, \ and\ \bibinfo {author}
  {\bibfnamefont {B.~O.}\ \bibnamefont {Wells}},\ }\href {\doibase
  10.1103/PhysRevLett.111.227001} {\bibfield  {journal} {\bibinfo  {journal}
  {Phys. Rev. Lett.}\ }\textbf {\bibinfo {volume} {111}},\ \bibinfo {pages}
  {227001} (\bibinfo {year} {2013})}\BibitemShut {NoStop}%
\bibitem [{\citenamefont {Xu}\ \emph {et~al.}(2014)\citenamefont {Xu},
  \citenamefont {Stock}, \citenamefont {Chi}, \citenamefont {Kolesnikov},
  \citenamefont {Xu}, \citenamefont {Gu},\ and\ \citenamefont
  {Tranquada}}]{Xu2014}%
  \BibitemOpen
  \bibfield  {author} {\bibinfo {author} {\bibfnamefont {Z.}~\bibnamefont
  {Xu}}, \bibinfo {author} {\bibfnamefont {C.}~\bibnamefont {Stock}}, \bibinfo
  {author} {\bibfnamefont {S.}~\bibnamefont {Chi}}, \bibinfo {author}
  {\bibfnamefont {A.~I.}\ \bibnamefont {Kolesnikov}}, \bibinfo {author}
  {\bibfnamefont {G.}~\bibnamefont {Xu}}, \bibinfo {author} {\bibfnamefont
  {G.}~\bibnamefont {Gu}}, \ and\ \bibinfo {author} {\bibfnamefont {J.~M.}\
  \bibnamefont {Tranquada}},\ }\href {\doibase 10.1103/PhysRevLett.113.177002}
  {\bibfield  {journal} {\bibinfo  {journal} {Phys. Rev. Lett.}\ }\textbf
  {\bibinfo {volume} {113}},\ \bibinfo {pages} {177002} (\bibinfo {year}
  {2014})}\BibitemShut {NoStop}%
\bibitem [{\citenamefont {Wagman}\ \emph {et~al.}(2016)\citenamefont {Wagman},
  \citenamefont {Carlo}, \citenamefont {Gaudet}, \citenamefont {Van~Gastel},
  \citenamefont {Abernathy}, \citenamefont {Stone}, \citenamefont {Granroth},
  \citenamefont {Kolesnikov}, \citenamefont {Savici}, \citenamefont {Kim},
  \citenamefont {Zhang}, \citenamefont {Ellis}, \citenamefont {Zhao},
  \citenamefont {Clark}, \citenamefont {Kallin}, \citenamefont {Mazurek},
  \citenamefont {Dabkowska},\ and\ \citenamefont {Gaulin}}]{Wagman2016}%
  \BibitemOpen
  \bibfield  {author} {\bibinfo {author} {\bibfnamefont {J.~J.}\ \bibnamefont
  {Wagman}}, \bibinfo {author} {\bibfnamefont {J.~P.}\ \bibnamefont {Carlo}},
  \bibinfo {author} {\bibfnamefont {J.}~\bibnamefont {Gaudet}}, \bibinfo
  {author} {\bibfnamefont {G.}~\bibnamefont {Van~Gastel}}, \bibinfo {author}
  {\bibfnamefont {D.~L.}\ \bibnamefont {Abernathy}}, \bibinfo {author}
  {\bibfnamefont {M.~B.}\ \bibnamefont {Stone}}, \bibinfo {author}
  {\bibfnamefont {G.~E.}\ \bibnamefont {Granroth}}, \bibinfo {author}
  {\bibfnamefont {A.~I.}\ \bibnamefont {Kolesnikov}}, \bibinfo {author}
  {\bibfnamefont {A.~T.}\ \bibnamefont {Savici}}, \bibinfo {author}
  {\bibfnamefont {Y.~J.}\ \bibnamefont {Kim}}, \bibinfo {author} {\bibfnamefont
  {H.}~\bibnamefont {Zhang}}, \bibinfo {author} {\bibfnamefont
  {D.}~\bibnamefont {Ellis}}, \bibinfo {author} {\bibfnamefont
  {Y.}~\bibnamefont {Zhao}}, \bibinfo {author} {\bibfnamefont {L.}~\bibnamefont
  {Clark}}, \bibinfo {author} {\bibfnamefont {A.~B.}\ \bibnamefont {Kallin}},
  \bibinfo {author} {\bibfnamefont {E.}~\bibnamefont {Mazurek}}, \bibinfo
  {author} {\bibfnamefont {H.~A.}\ \bibnamefont {Dabkowska}}, \ and\ \bibinfo
  {author} {\bibfnamefont {B.~D.}\ \bibnamefont {Gaulin}},\ }\href {\doibase
  10.1103/PhysRevB.93.094416} {\bibfield  {journal} {\bibinfo  {journal} {Phys.
  Rev. B}\ }\textbf {\bibinfo {volume} {93}},\ \bibinfo {pages} {094416}
  (\bibinfo {year} {2016})}\BibitemShut {NoStop}%
\bibitem [{\citenamefont {R{\o}mer}\ \emph {et~al.}(2013)\citenamefont
  {R{\o}mer}, \citenamefont {Chang}, \citenamefont {Christensen}, \citenamefont
  {Andersen}, \citenamefont {Lefmann}, \citenamefont {M{\"a}hler},
  \citenamefont {Gavilano}, \citenamefont {Gilardi}, \citenamefont
  {Niedermayer}, \citenamefont {R{\o}nnow}, \citenamefont {Schneidewind},
  \citenamefont {Link}, \citenamefont {Oda}, \citenamefont {Ido}, \citenamefont
  {Momono},\ and\ \citenamefont {Mesot}}]{romer2013glassy}%
  \BibitemOpen
  \bibfield  {author} {\bibinfo {author} {\bibfnamefont {A.~T.}\ \bibnamefont
  {R{\o}mer}}, \bibinfo {author} {\bibfnamefont {J.}~\bibnamefont {Chang}},
  \bibinfo {author} {\bibfnamefont {N.~B.}\ \bibnamefont {Christensen}},
  \bibinfo {author} {\bibfnamefont {B.~M.}\ \bibnamefont {Andersen}}, \bibinfo
  {author} {\bibfnamefont {K.}~\bibnamefont {Lefmann}}, \bibinfo {author}
  {\bibfnamefont {L.}~\bibnamefont {M{\"a}hler}}, \bibinfo {author}
  {\bibfnamefont {J.}~\bibnamefont {Gavilano}}, \bibinfo {author}
  {\bibfnamefont {R.}~\bibnamefont {Gilardi}}, \bibinfo {author} {\bibfnamefont
  {C.}~\bibnamefont {Niedermayer}}, \bibinfo {author} {\bibfnamefont {H.~M.}\
  \bibnamefont {R{\o}nnow}}, \bibinfo {author} {\bibfnamefont {A.}~\bibnamefont
  {Schneidewind}}, \bibinfo {author} {\bibfnamefont {P.}~\bibnamefont {Link}},
  \bibinfo {author} {\bibfnamefont {M.}~\bibnamefont {Oda}}, \bibinfo {author}
  {\bibfnamefont {M.}~\bibnamefont {Ido}}, \bibinfo {author} {\bibfnamefont
  {N.}~\bibnamefont {Momono}}, \ and\ \bibinfo {author} {\bibfnamefont
  {J.}~\bibnamefont {Mesot}},\ }\href@noop {} {\bibfield  {journal} {\bibinfo
  {journal} {Physical Review B}\ }\textbf {\bibinfo {volume} {87}},\ \bibinfo
  {pages} {144513} (\bibinfo {year} {2013})}\BibitemShut {NoStop}%
\bibitem [{\citenamefont {Singer}\ \emph {et~al.}(2002)\citenamefont {Singer},
  \citenamefont {Hunt},\ and\ \citenamefont {Imai}}]{sing02}%
  \BibitemOpen
  \bibfield  {author} {\bibinfo {author} {\bibfnamefont {P.~M.}\ \bibnamefont
  {Singer}}, \bibinfo {author} {\bibfnamefont {A.~W.}\ \bibnamefont {Hunt}}, \
  and\ \bibinfo {author} {\bibfnamefont {T.}~\bibnamefont {Imai}},\ }\href
  {\doibase 10.1103/PhysRevLett.88.047602} {\bibfield  {journal} {\bibinfo
  {journal} {Phys. Rev. Lett.}\ }\textbf {\bibinfo {volume} {88}},\ \bibinfo
  {pages} {047602} (\bibinfo {year} {2002})}\BibitemShut {NoStop}%
\bibitem [{\citenamefont {Singer}\ \emph {et~al.}(2005)\citenamefont {Singer},
  \citenamefont {Imai}, \citenamefont {Chou}, \citenamefont {Hirota},
  \citenamefont {Takaba}, \citenamefont {Kakeshita}, \citenamefont {Eisaki},\
  and\ \citenamefont {Uchida}}]{sing05}%
  \BibitemOpen
  \bibfield  {author} {\bibinfo {author} {\bibfnamefont {P.~M.}\ \bibnamefont
  {Singer}}, \bibinfo {author} {\bibfnamefont {T.}~\bibnamefont {Imai}},
  \bibinfo {author} {\bibfnamefont {F.~C.}\ \bibnamefont {Chou}}, \bibinfo
  {author} {\bibfnamefont {K.}~\bibnamefont {Hirota}}, \bibinfo {author}
  {\bibfnamefont {M.}~\bibnamefont {Takaba}}, \bibinfo {author} {\bibfnamefont
  {T.}~\bibnamefont {Kakeshita}}, \bibinfo {author} {\bibfnamefont
  {H.}~\bibnamefont {Eisaki}}, \ and\ \bibinfo {author} {\bibfnamefont
  {S.}~\bibnamefont {Uchida}},\ }\href {\doibase 10.1103/PhysRevB.72.014537}
  {\bibfield  {journal} {\bibinfo  {journal} {Phys. Rev. B}\ }\textbf {\bibinfo
  {volume} {72}},\ \bibinfo {pages} {014537} (\bibinfo {year}
  {2005})}\BibitemShut {NoStop}%
\bibitem [{\citenamefont {Jacobsen}\ \emph {et~al.}(2018)\citenamefont
  {Jacobsen}, \citenamefont {Holm}, \citenamefont {L\ifmmode \u{a}\else
  \u{a}\fi{}c\ifmmode \u{a}\else \u{a}\fi{}tu\ifmmode~\mbox{\c{s}}\else
  \c{s}\fi{}u}, \citenamefont {R\o{}mer}, \citenamefont {Bertelsen},
  \citenamefont {Boehm}, \citenamefont {Toft-Petersen}, \citenamefont {Grivel},
  \citenamefont {Emery}, \citenamefont {Udby}, \citenamefont {Wells},\ and\
  \citenamefont {Lefmann}}]{Jacobsen2018}%
  \BibitemOpen
  \bibfield  {author} {\bibinfo {author} {\bibfnamefont {H.}~\bibnamefont
  {Jacobsen}}, \bibinfo {author} {\bibfnamefont {S.~L.}\ \bibnamefont {Holm}},
  \bibinfo {author} {\bibfnamefont {M.-E.}\ \bibnamefont {L\ifmmode \u{a}\else
  \u{a}\fi{}c\ifmmode \u{a}\else \u{a}\fi{}tu\ifmmode~\mbox{\c{s}}\else
  \c{s}\fi{}u}}, \bibinfo {author} {\bibfnamefont {A.~T.}\ \bibnamefont
  {R\o{}mer}}, \bibinfo {author} {\bibfnamefont {M.}~\bibnamefont {Bertelsen}},
  \bibinfo {author} {\bibfnamefont {M.}~\bibnamefont {Boehm}}, \bibinfo
  {author} {\bibfnamefont {R.}~\bibnamefont {Toft-Petersen}}, \bibinfo {author}
  {\bibfnamefont {J.-C.}\ \bibnamefont {Grivel}}, \bibinfo {author}
  {\bibfnamefont {S.~B.}\ \bibnamefont {Emery}}, \bibinfo {author}
  {\bibfnamefont {L.}~\bibnamefont {Udby}}, \bibinfo {author} {\bibfnamefont
  {B.~O.}\ \bibnamefont {Wells}}, \ and\ \bibinfo {author} {\bibfnamefont
  {K.}~\bibnamefont {Lefmann}},\ }\href {\doibase
  10.1103/PhysRevLett.120.037003} {\bibfield  {journal} {\bibinfo  {journal}
  {Phys. Rev. Lett.}\ }\textbf {\bibinfo {volume} {120}},\ \bibinfo {pages}
  {037003} (\bibinfo {year} {2018})}\BibitemShut {NoStop}%
\end{thebibliography}%


\begin{thebibliography}{34}%
\makeatletter
\providecommand \@ifxundefined [1]{%
 \@ifx{#1\undefined}
}%
\providecommand \@ifnum [1]{%
 \ifnum #1\expandafter \@firstoftwo
 \else \expandafter \@secondoftwo
 \fi
}%
\providecommand \@ifx [1]{%
 \ifx #1\expandafter \@firstoftwo
 \else \expandafter \@secondoftwo
 \fi
}%
\providecommand \natexlab [1]{#1}%
\providecommand \enquote  [1]{``#1''}%
\providecommand \bibnamefont  [1]{#1}%
\providecommand \bibfnamefont [1]{#1}%
\providecommand \citenamefont [1]{#1}%
\providecommand \href@noop [0]{\@secondoftwo}%
\providecommand \href [0]{\begingroup \@sanitize@url \@href}%
\providecommand \@href[1]{\@@startlink{#1}\@@href}%
\providecommand \@@href[1]{\endgroup#1\@@endlink}%
\providecommand \@sanitize@url [0]{\catcode `\\12\catcode `\$12\catcode
  `\&12\catcode `\#12\catcode `\^12\catcode `\_12\catcode `\%12\relax}%
\providecommand \@@startlink[1]{}%
\providecommand \@@endlink[0]{}%
\providecommand \url  [0]{\begingroup\@sanitize@url \@url }%
\providecommand \@url [1]{\endgroup\@href {#1}{\urlprefix }}%
\providecommand \urlprefix  [0]{URL }%
\providecommand \Eprint [0]{\href }%
\providecommand \doibase [0]{http://dx.doi.org/}%
\providecommand \selectlanguage [0]{\@gobble}%
\providecommand \bibinfo  [0]{\@secondoftwo}%
\providecommand \bibfield  [0]{\@secondoftwo}%
\providecommand \translation [1]{[#1]}%
\providecommand \BibitemOpen [0]{}%
\providecommand \bibitemStop [0]{}%
\providecommand \bibitemNoStop [0]{.\EOS\space}%
\providecommand \EOS [0]{\spacefactor3000\relax}%
\providecommand \BibitemShut  [1]{\csname bibitem#1\endcsname}%
\let\auto@bib@innerbib\@empty
\bibitem [{\citenamefont {Tanaka}\ \emph {et~al.}(1989)\citenamefont {Tanaka},
  \citenamefont {Yamane},\ and\ \citenamefont {Kojima}}]{Tanaka89}%
  \BibitemOpen
  \bibfield  {author} {\bibinfo {author} {\bibfnamefont {I.}~\bibnamefont
  {Tanaka}}, \bibinfo {author} {\bibfnamefont {K.}~\bibnamefont {Yamane}}, \
  and\ \bibinfo {author} {\bibfnamefont {H.}~\bibnamefont {Kojima}},\ }\href
  {\doibase https://doi.org/10.1016/0022-0248(89)90074-2} {\bibfield  {journal}
  {\bibinfo  {journal} {Journal of Crystal Growth}\ }\textbf {\bibinfo {volume}
  {96}},\ \bibinfo {pages} {711} (\bibinfo {year} {1989})}\BibitemShut
  {NoStop}%
\bibitem [{\citenamefont {Boehm}\ \emph {et~al.}(2015)\citenamefont {Boehm},
  \citenamefont {Steffens}, \citenamefont {Kulda}, \citenamefont {Klicpera},
  \citenamefont {Roux}, \citenamefont {Courtois}, \citenamefont {Svoboda},
  \citenamefont {Saroun},\ and\ \citenamefont {Sechovsky}}]{boehm2015thales}%
  \BibitemOpen
  \bibfield  {author} {\bibinfo {author} {\bibfnamefont {M.}~\bibnamefont
  {Boehm}}, \bibinfo {author} {\bibfnamefont {P.}~\bibnamefont {Steffens}},
  \bibinfo {author} {\bibfnamefont {J.}~\bibnamefont {Kulda}}, \bibinfo
  {author} {\bibfnamefont {M.}~\bibnamefont {Klicpera}}, \bibinfo {author}
  {\bibfnamefont {S.}~\bibnamefont {Roux}}, \bibinfo {author} {\bibfnamefont
  {P.}~\bibnamefont {Courtois}}, \bibinfo {author} {\bibfnamefont
  {P.}~\bibnamefont {Svoboda}}, \bibinfo {author} {\bibfnamefont
  {J.}~\bibnamefont {Saroun}}, \ and\ \bibinfo {author} {\bibfnamefont
  {V.}~\bibnamefont {Sechovsky}},\ }\href@noop {} {\bibfield  {journal}
  {\bibinfo  {journal} {Neutron News}\ }\textbf {\bibinfo {volume} {26}},\
  \bibinfo {pages} {18} (\bibinfo {year} {2015})}\BibitemShut {NoStop}%
\bibitem [{\citenamefont {Sandberg~{\em et al.}}(2022)}]{Sandberg2022}%
  \BibitemOpen
  \bibfield  {author} {\bibinfo {author} {\bibfnamefont {L.~{\O}.}\
  \bibnamefont {Sandberg~{\em et al.}}},\ }\href@noop {} {\bibfield  {journal}
  {\bibinfo  {journal} {manuscript in preparation}\ } (\bibinfo {year}
  {2022})}\BibitemShut {NoStop}%
\bibitem [{\citenamefont {R{\o}mer}\ \emph {et~al.}(2015)\citenamefont
  {R{\o}mer}, \citenamefont {Ray}, \citenamefont {Jacobsen}, \citenamefont
  {Udby}, \citenamefont {Andersen}, \citenamefont {Bertelsen}, \citenamefont
  {Holm}, \citenamefont {Christensen}, \citenamefont {Toft-Petersen},
  \citenamefont {Skoulatos}, \citenamefont {Laver}, \citenamefont
  {Schneidewind}, \citenamefont {Link}, \citenamefont {Oda}, \citenamefont
  {Ido}, \citenamefont {Momono},\ and\ \citenamefont {Lefmann}}]{Roemer2015}%
  \BibitemOpen
  \bibfield  {author} {\bibinfo {author} {\bibfnamefont {A.~T.}\ \bibnamefont
  {R{\o}mer}}, \bibinfo {author} {\bibfnamefont {P.~J.}\ \bibnamefont {Ray}},
  \bibinfo {author} {\bibfnamefont {H.}~\bibnamefont {Jacobsen}}, \bibinfo
  {author} {\bibfnamefont {L.}~\bibnamefont {Udby}}, \bibinfo {author}
  {\bibfnamefont {B.~M.}\ \bibnamefont {Andersen}}, \bibinfo {author}
  {\bibfnamefont {M.}~\bibnamefont {Bertelsen}}, \bibinfo {author}
  {\bibfnamefont {S.~L.}\ \bibnamefont {Holm}}, \bibinfo {author}
  {\bibfnamefont {N.~B.}\ \bibnamefont {Christensen}}, \bibinfo {author}
  {\bibfnamefont {R.}~\bibnamefont {Toft-Petersen}}, \bibinfo {author}
  {\bibfnamefont {M.}~\bibnamefont {Skoulatos}}, \bibinfo {author}
  {\bibfnamefont {M.}~\bibnamefont {Laver}}, \bibinfo {author} {\bibfnamefont
  {A.}~\bibnamefont {Schneidewind}}, \bibinfo {author} {\bibfnamefont
  {P.}~\bibnamefont {Link}}, \bibinfo {author} {\bibfnamefont {M.}~\bibnamefont
  {Oda}}, \bibinfo {author} {\bibfnamefont {M.}~\bibnamefont {Ido}}, \bibinfo
  {author} {\bibfnamefont {N.}~\bibnamefont {Momono}}, \ and\ \bibinfo {author}
  {\bibfnamefont {K.}~\bibnamefont {Lefmann}},\ }\href@noop {} {\bibfield
  {journal} {\bibinfo  {journal} {Physical Review B}\ }\textbf {\bibinfo
  {volume} {91}},\ \bibinfo {pages} {174507} (\bibinfo {year}
  {2015})}\BibitemShut {NoStop}%
\bibitem [{\citenamefont {Tranquada}\ \emph {et~al.}(2008)\citenamefont
  {Tranquada}, \citenamefont {Gu}, \citenamefont {H{\"u}cker}, \citenamefont
  {Jie}, \citenamefont {Kang}, \citenamefont {Klingeler}, \citenamefont {Li},
  \citenamefont {Tristan}, \citenamefont {Wen}, \citenamefont {Xu},
  \citenamefont {Xu}, \citenamefont {Zhou},\ and\ \citenamefont
  {Zimmermann}}]{tranquada2008evidence}%
  \BibitemOpen
  \bibfield  {author} {\bibinfo {author} {\bibfnamefont {J.~M.}\ \bibnamefont
  {Tranquada}}, \bibinfo {author} {\bibfnamefont {G.~D.}\ \bibnamefont {Gu}},
  \bibinfo {author} {\bibfnamefont {M.}~\bibnamefont {H{\"u}cker}}, \bibinfo
  {author} {\bibfnamefont {Q.}~\bibnamefont {Jie}}, \bibinfo {author}
  {\bibfnamefont {H.-J.}\ \bibnamefont {Kang}}, \bibinfo {author}
  {\bibfnamefont {R.}~\bibnamefont {Klingeler}}, \bibinfo {author}
  {\bibfnamefont {Q.}~\bibnamefont {Li}}, \bibinfo {author} {\bibfnamefont
  {N.}~\bibnamefont {Tristan}}, \bibinfo {author} {\bibfnamefont {J.~S.}\
  \bibnamefont {Wen}}, \bibinfo {author} {\bibfnamefont {G.~Y.}\ \bibnamefont
  {Xu}}, \bibinfo {author} {\bibfnamefont {Z.~J.}\ \bibnamefont {Xu}}, \bibinfo
  {author} {\bibfnamefont {J.}~\bibnamefont {Zhou}}, \ and\ \bibinfo {author}
  {\bibfnamefont {M.~v.}\ \bibnamefont {Zimmermann}},\ }\href@noop {}
  {\bibfield  {journal} {\bibinfo  {journal} {Physical Review B}\ }\textbf
  {\bibinfo {volume} {78}},\ \bibinfo {pages} {174529} (\bibinfo {year}
  {2008})}\BibitemShut {NoStop}%
\end{thebibliography}%




\end{document}


\title{Supplemental Material: \\ Evolution of magnetic stripes under uniaxial stress in \ce{La_{1.885}Ba_{0.115}CuO4} studied by neutron scattering}

\author{Machteld E. Kamminga}
\email{kamminga@nbi.ku.dk}
\affiliation{Nanoscience Center, Niels Bohr Institute, University of Copenhagen, 2100 Copenhagen {\O}, Denmark}
\author{Kristine M. L. Krighaar}
\author{Astrid T. R{\o}mer}
\author{Lise {\O}. Sandberg}
\affiliation{Nanoscience Center, Niels Bohr Institute, University of Copenhagen, 2100 Copenhagen {\O}, Denmark}
\author{Pascale P. Deen}
\affiliation{Nanoscience Center, Niels Bohr Institute, University of Copenhagen, 2100 Copenhagen {\O}, Denmark}
\affiliation{European Spallation Source ERIC, Partikelgatan 224, 84 Lund, Sweden}
\author{Martin Boehm}
\affiliation{Institut Laue-Langevin, 71 Avenue des Martyrs, CS 20156, 38042 Grenoble Cedex 9,
France}
\author{G. D. Gu}
\author{J. M. Tranquada}
\affiliation{Condensed Matter Physics and Materials Science Division, Brookhaven National Laboratory,
Upton, New York 11973, USA}
\author{Kim Lefmann}
\email{lefmann@nbi.ku.dk}
\affiliation{Nanoscience Center, Niels Bohr Institute, University of Copenhagen, 2100 Copenhagen {\O}, Denmark}

\maketitle
\section{Experimental details}
\noindent The sample is a 0.76~g single crystal of LBCO, $x=0.115$  with $T_{\rm c} \sim 12$~K, grown by the travelling-solvent floating-zone method.\cite{Tanaka89} The sample is cylindrically shaped with 5~mm diameter and a height of 3.9~mm. To determine the orientation of the lattice, white-beam X-ray Laue diffraction from an Ag-anode showed that the $c$-axis lies almost perfectly perpendicular to the cylinder axis, while one of the other main symmetry axes lies parallel to the cylinder axis. To determine the true orientation, neutron diffraction was performed at the two axis diffractometer ORION at PSI. The tetragonal [1$\overline{1}$0] direction was found to be along the cylinder axis, and the scan showed a rocking curve of the (200) peak gave a peak width of 0.54 degrees (FWHM) as an upper limit for the sample mosaicity.

The sample was polished to assure that the cylinder surfaces were as parallel as possible, in order to hinder the crystal from breaking under applied stress on the cylinder surfaces. The polishing was performed with an in-house custom-build  grinder.  

The main experiment took place on the cold neutron triple-axis spectrometer ThALES at Institut Laue-Langevin (ILL), Grenoble.\cite{boehm2015thales}  All the measurements were performed with a constant $E_f = 5$~meV and used a double focusing monochromator and a horizontally focusing analyzer. The sample was oriented with the cylinder axis vertical and the [110] and [001] directions in the scattering plane. 

The crystal was subjected to uniaxial stress along the tetragonal [1$\overline{1}$0] \textit{i.e.}\ along the diagonal Cu-Cu direction, using a home-built pressure cell of 46~mm diameter, optimized for neutron scattering.\cite{Sandberg2022} The pressure sensor was calibrated off-site at room temperature and at cryogenic temperatures.\cite{Sandberg2022} 

We performed comparable elastic and inelastic measurements, with and without applied stress. First measurements were performed at ambient pressure, where the sample was placed inside the unloaded pressure cell with only the weight of the 180~g piston and steel seats, fixing the sample in the neutron beam. With this set-up, the pressure readout in the fully assembled cell using the in-situ monitor showed 7.4~MPa at room temperature and 1.7~MPa at the cryostat base temperature. These values falls well within the errors of the calibration readout, \cite{Sandberg2022} and we are therefore confident that the pressure from the piston is negligible and we are in fact at zero pressure. We therefore use the reading for the unloaded cell as the zero point for the pressure monitor. 

The pressure cell and sample was cooled in a standard Orange cryostat with a 50~mm bore, modified for the cabling of the cell. We first performed alignment of the crystal orientation using the (110) and (002) structural Bragg peaks. The scans in reciprocal space for the elastic and inelastic magnetic stripe signals were performed as sample rotation scans, in order to keep the background constant.\cite{Roemer2015} First the inelastic signal was measured at the incommensurate stripe position at (0.615 0.5 2) at energy transfer 1.0~meV and 20~K, using no beam collimation. Afterwards, $40'$ collimation was inserted before and after the sample to improve resolution and reduce the background to measure the elastic signal at 2~K, 20~K, 45~K at the (0.615 0.5 2) incommensurate position. The inelastic measurements were taken at approximately 46 min per point while the elastics were taken at 14 min per point.

Halfway through the experiment, the sample stick was taken out, as stress had to be applied manually. The stress was applied slowly by 4 top screws until a stress of 30~MPa at room temperature was achieved, as judged by the pressure read-out.  
During the re-insertion of the sample stick, the copper wires allowing for pressure readout of the cell became disconnected.
The wires were attempted to be re-soldered but unsuccessfully, preventing in-situ measurements of the pressure while in the neutron beam. After applying stress, we re-measured the (110) and (002) Bragg peaks, confirming that the crystal was still aligned and in one piece. We choose a slow cooling rate of of 0.5 K per minute from 300 K to 200 K to diminish the thermal shock when the crystal goes from the high temperature orthogonal to a low temperature tetragonal phase. 

When aligning we found that the stress had changed the lattice parameters and they were adjusted accordingly in software.\cite{Sandberg2022}  We measured the elastic signal again at 2~K and 20~K, with the $40'$ collimation unchanged. Due to time constraints, the background measurements at 45~K were not repeated, as we did not expect any change with stress. Subsequently, the collimation was taken out and we measured the inelastic stripe signal with applied stress at 20~K. For the inelastic scans without applied stress, 27e6 monitor counts were taken, while for the ones with applied stress, 26e6 monitor counts were registered due to time constraints. However, the data were re-scaled accordingly.

After applying stress and reinserting the cell into the cryostat, a change in background was observed. We argue that this change could be caused by a small change in cell orientation, especially because the change in magnetic signal at 2 K is only half of that of the change in background (Fig.~1 in the main text), and the magnetic stripes only fill a tiny fraction of the Brillouin zone, whereas the background is everywhere in reciprocal space. As a result, the integrated change in background signal is around three orders of magnitude larger than the change in magnetic signal, implying that the changes in magnetic signal and background are not correlated. With this insight, a background subtraction was performed on the raw data, displayed as Fig.~1 and Fig.~3 in the main text, for easy comparison. 

As a last action, we measured the transverse acoustic phonon around the (110) position at 2.0~meV and at 90~K. 

Since we were unable to obtain a pressure read-out during the experiment, the sample stayed in the cell at room temperature for 10~days, until the wires could be re-soldered. The stress was then measured to be 15~GPa. We know that the pressure cell is prone to lose pressure at room temperature, but maintains very well at slightly lower temperatures, {\em e.g.}~at 250~K.\cite{Sandberg2022} Therefore, we can assume that the stress of 30~GPa was maintained during at the actual neutron experiment.

\section{Scattering geometry}
\noindent The sample was nominally aligned with the [110] and [001] directions in the scattering plane.
Initial alignment on the Bragg peaks showed that the [1$\overline{1}$0] crystal direction was $3.0^\circ$ away from the scattering plane when the sample cell was vertical, {\em i.e.}\ that the stress was applied 3.0$^\circ$ off the desired [1$\overline{1}$0] direction. The misalignment of the [001] direction was much smaller, around $1^\circ$, which was corrected at the corresponding goniometer.

For our measurements, we aimed to access the (0.615 0.5 $L$) position in reciprocal space. To fulfill this in the present setting, we rotated the crystal, so that the [1$\overline{1}$0] direction pointed $5.9^\circ$ out of the scattering plane. This rotation was chosen to be in the opposite direction from the alignment error, so that in the final experiment the sample cell pointed $2.9^\circ$ off the vertical direction.

For the first inelastic measurements at zero stress, we accidentally centered the sample rotation scan around the position (0.625 0.508 2.000), in stead of the correct value of (0.615 0.500 2.000). For consistency, we kept this slightly wrong center value also at the in-stress inelastic scan. Due to the rod-like nature of the stripes, however, this mishap had negligible effect on the experiment. Simple trigonometry shows that the sample rotation scan would intersect the stripe signal at the position (0.615 0.500 2.038), which has an $L$ value so close to the wanted value of 2.000 that the difference in signal is almost non-existing. The same geometrical effect ensures that even if the inelastic stripe signal moved under stress, we can still confidently say that we have sampled the signal in comparable ways in the two scans.

A very similar alignment procedure was performed by us in an earlier experiment on \ce{La_{2-x}Sr_xCuO4}, where we also aimed to have the [001] direction in the scattering plane \cite{Roemer2015}. Here we found that the elastic stripe signal is very slowly modulated, as a function of $L$, almost sinusoidal with a period of 2.

\section{Raw data and data fitting}
\begin{figure}[h!]
\includegraphics[width=0.35\textwidth]{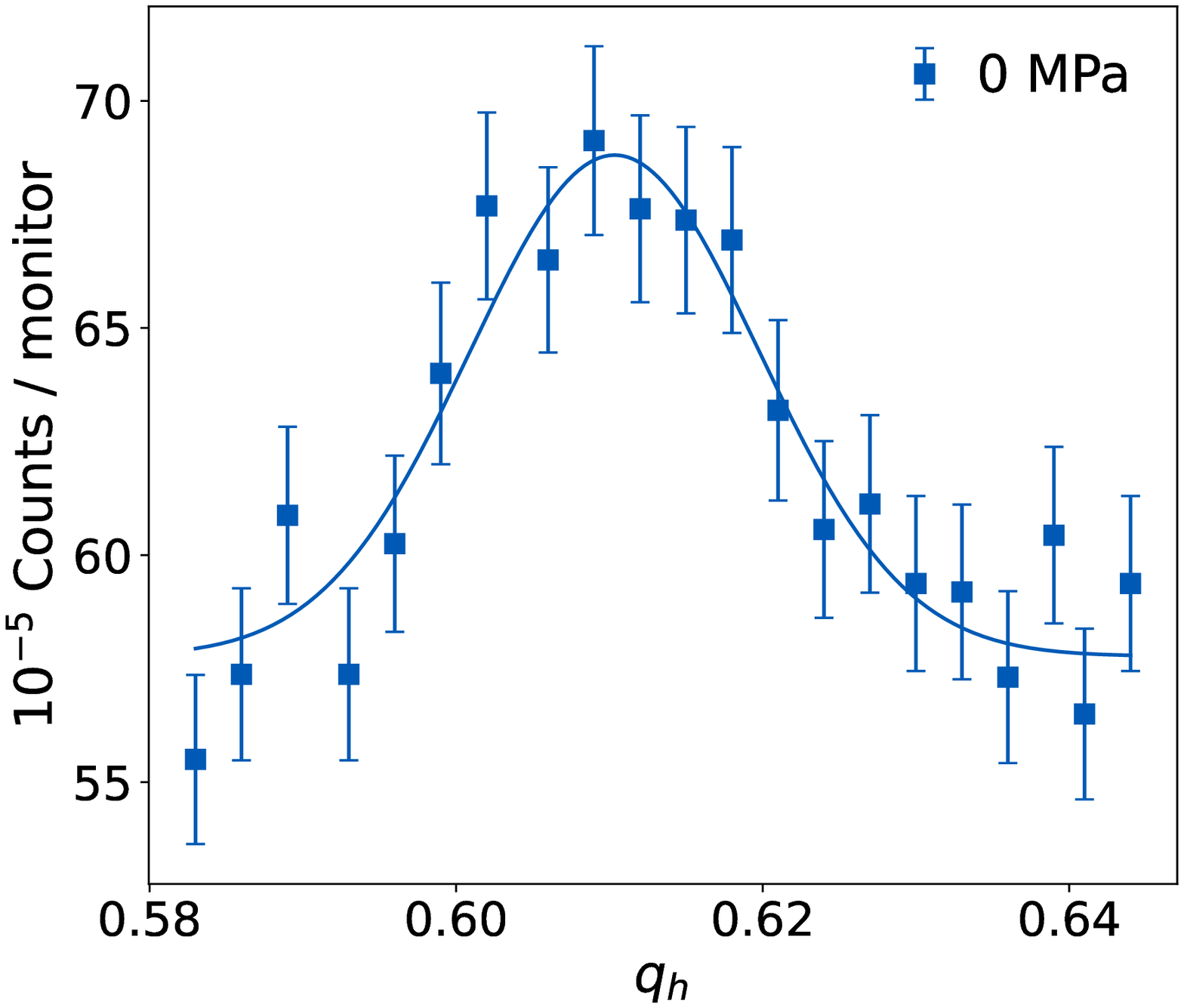}
\caption{Summation of the elastic signals at 2 and 20 K without applied stress and without background subtraction, used to determine the peak width that is used as a fixed parameter in Figs.~\ref{fig:2-20elastic} and \ref{fig:45K}. The fitting parameters are given in Table \ref{Tab:fitparameters0MPasummed}.}
\label{Fig:0MPasummed}
\end{figure}

\noindent We first show the sum of the elastic scans at ambient pressure taken at 2~K and 20~K, see Fig.~\ref{Fig:0MPasummed}. We use the sum of these two data sets to improve statistics, in order to better determine the fitting parameters, see Table~\ref{Tab:fitparameters0MPasummed}. We use this Gaussian width (sigma) as a fixed parameter for further data analysis, since all previous experiments on stripes have shown that the peak width is independent of temperature within the ordered phase.

\begin{table}[h!]
\begin{tabular}{|l|l|l|}
      \hline & \ 0 meV, 0 MPa \ \\ \hline
\ $a$ \ $(10^{-5})$ \ & \ 11.035 $\pm$ 3.319  \  \\
\ $mu$ \    & \ 0.610 $\pm$  0.001   \   \\
\ $sigma$ \ & \ 0.0095 $\pm$ 0.0014 \     \\
\ $bg$ \ $(10^{-5})$ \   & \ 57.772 $\pm$  0.826 \   \\ \hline
\end{tabular}
\caption{Fitting parameters of the Gaussian fit to the summed elastic signals shown in Fig.~\ref{Fig:0MPasummed}, with $a$ the height of the peak, $mu$ the peak position, $sigma$ the standard deviation and $bg$ the background. $a$ and $bg$ are given in counts per monitor.}
\label{Tab:fitparameters0MPasummed}
\end{table}

\begin{figure}[h!]
\centering
\includegraphics[width=.35\linewidth]{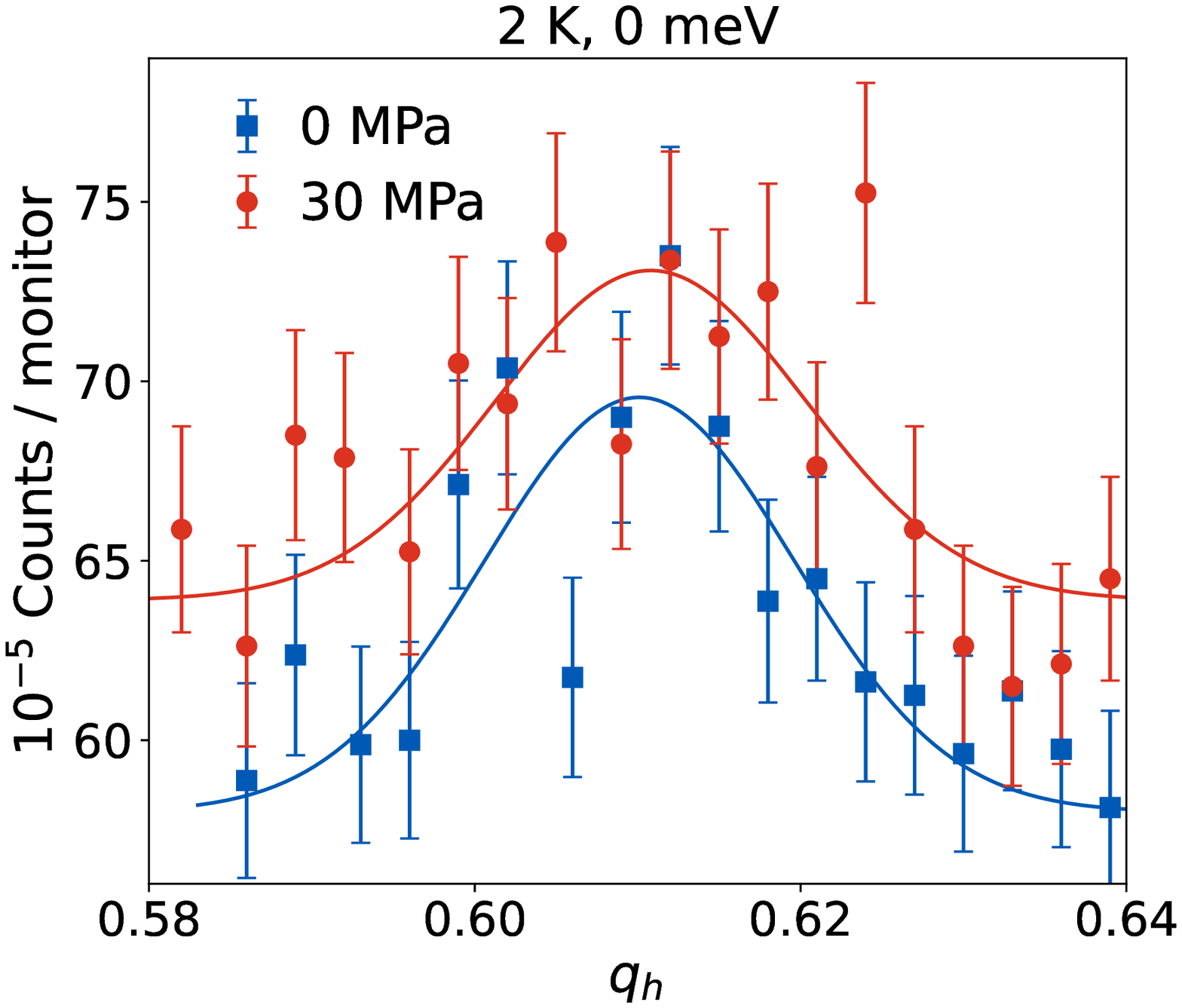}
\includegraphics[width=.35\linewidth]{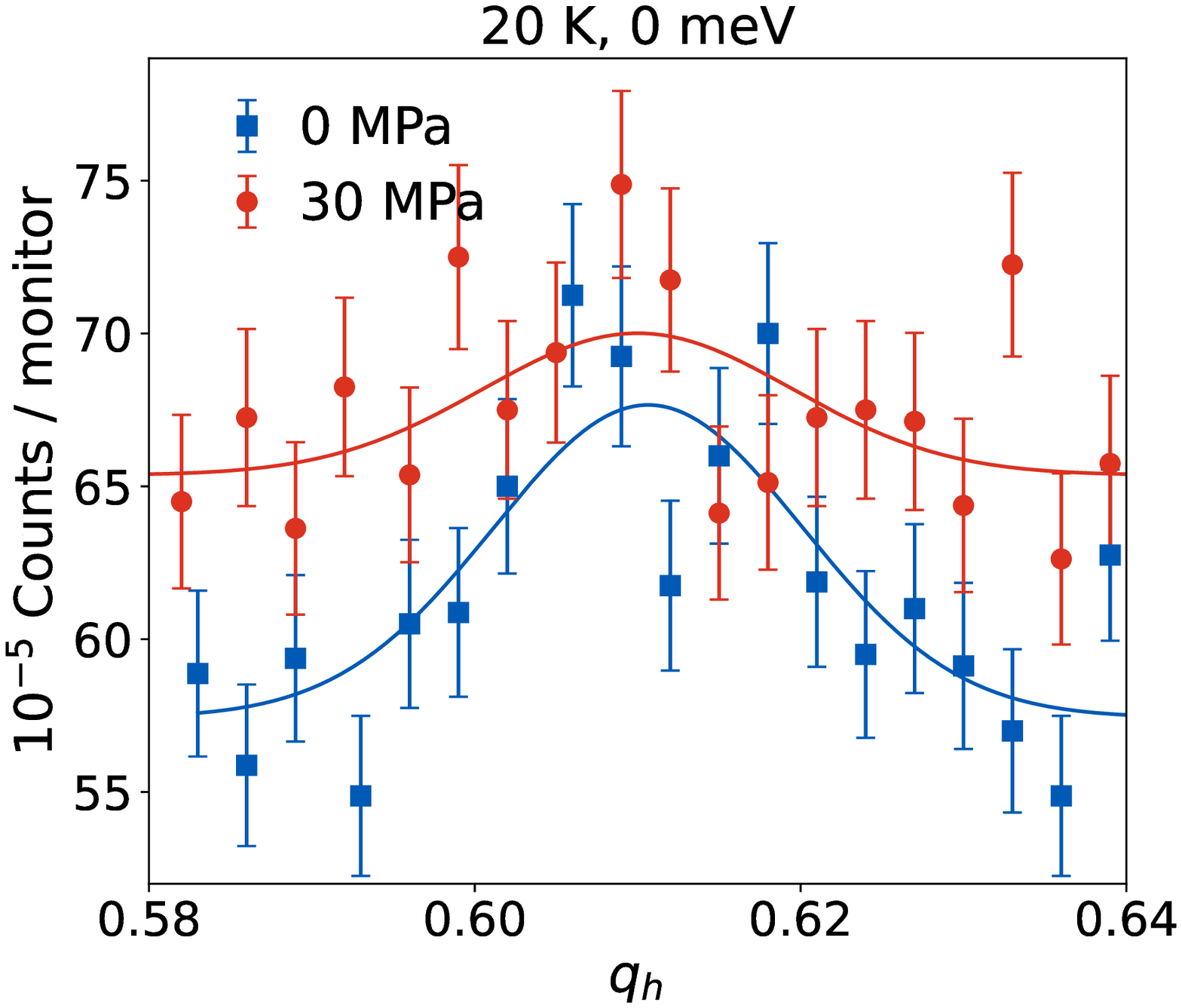}
\caption{Elastic signal at 2 K (left) and 20 K (right) without applied stress and with 30 MPa, without background subtraction. The fitting parameters are given in Table \ref{Tab:fitparameterselastic}.}
\label{fig:2-20elastic}
\end{figure}

\begin{figure}[h!]
\includegraphics[width=0.35\textwidth]{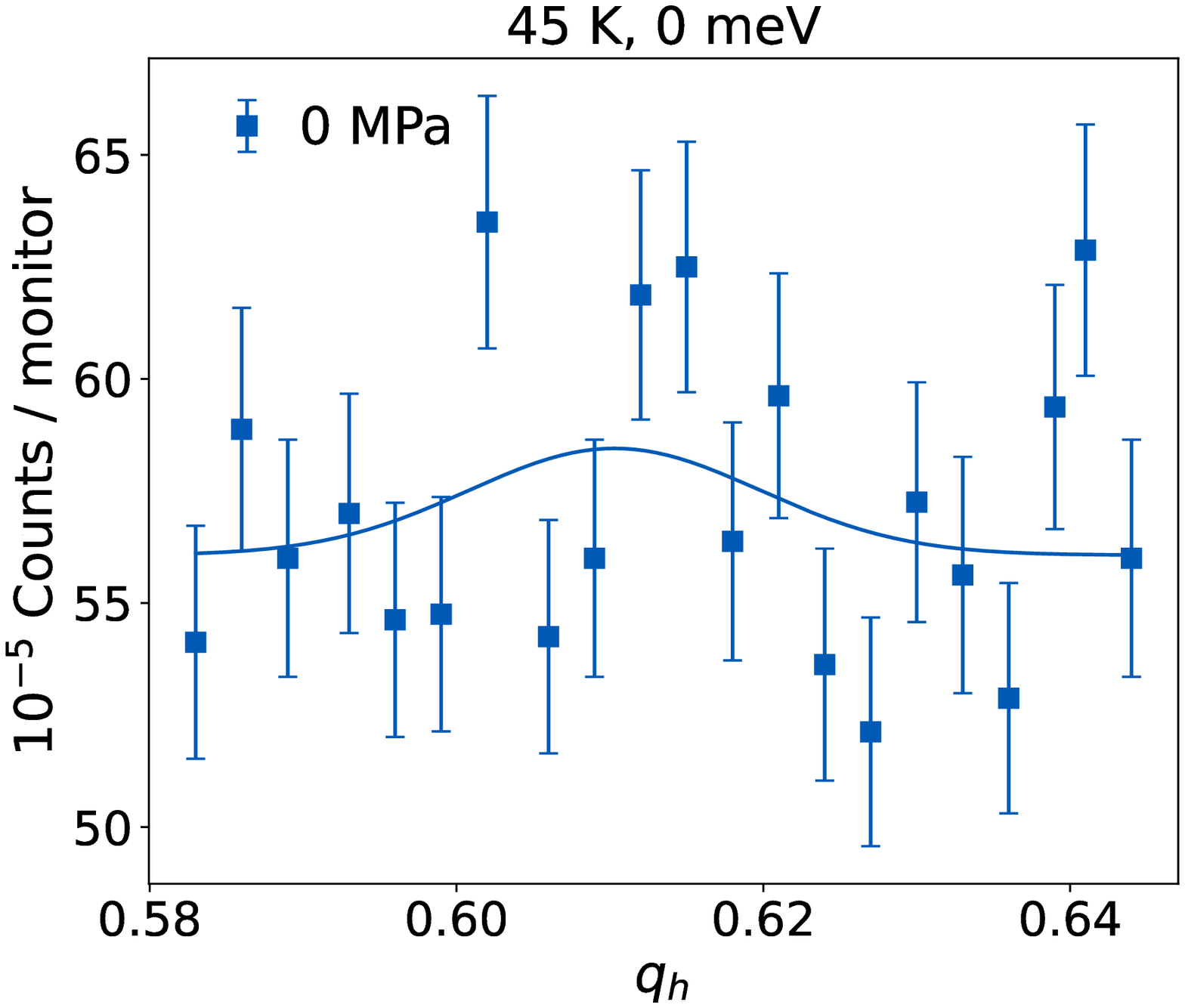}
\caption{Elastic signal at 45 K without applied stress and without background subtraction.}
\label{fig:45K}
\end{figure}

As a next step, we show all elastic data, taken with and without 30~MPa stress at temperatures of 2~K and 20~K, see Fig.~\ref{fig:2-20elastic}, and at the assumed background temperature of 45~K, see Fig.~\ref{fig:45K}. Note that a small elastic intensity was found at 45~K at 0~MPa, with the error bars nearly reaching zero. However, we argue that this residual intensity is caused by integrating over the instrumental resolution of 0.2~meV, which therefore incorporates a small fraction of the dynamic stripes that extend well beyond $T_{\rm c}$.\cite{tranquada2008evidence}

\begin{figure}[h!]
\includegraphics[width=0.35\textwidth]{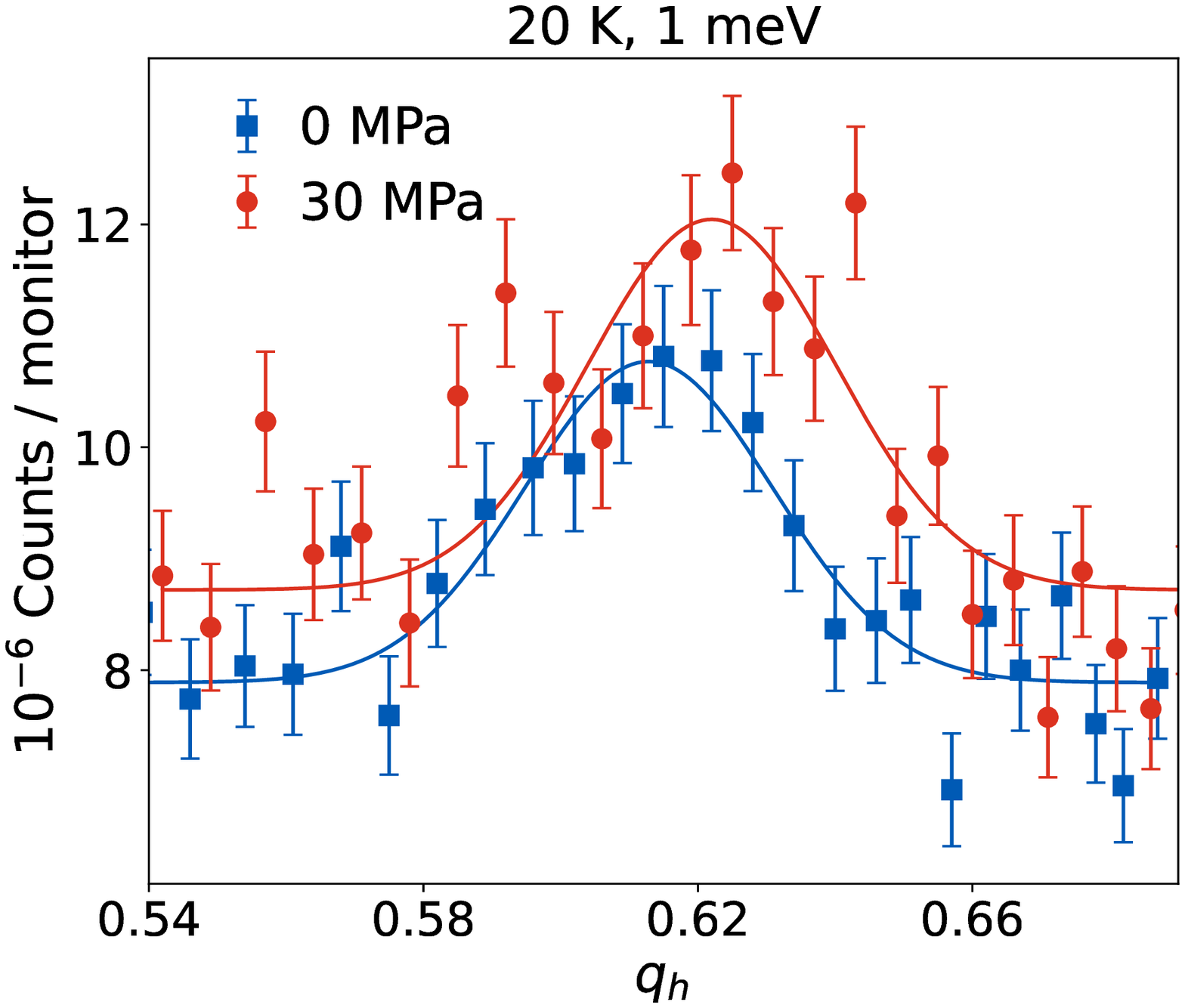}
\caption{Inelastic signals (1 meV) at 20 K without applied stress and with 30 MPa, without background subtraction. The fitting parameters are given in Table \ref{Tab:fitparametersinelastic}.}
\label{fig:inelastic}
\end{figure}

Finally, we show the raw inelastic data at 20~K, taken at zero and 30~MPa stress, see Fig.~\ref{fig:inelastic}. The corresponding fitting parameters are given in Table~\ref{Tab:fitparametersinelastic}. We notice a significant change between the positions of the two peaks, and a hint of an increase in amplitude for the pressure data. We note that since the sample is oriented with [001] and [0.615 0.5 0] directions in the scattering plane, a shift in the value of the $h$ component of a reciprocal point also leads to a similar shift in the respective $k$ value, and one could consider if the scan is traversing the maximum of the stripe signal, earlier reported to be positioned at (0.5+$\delta$ 0.5 $L$). For a shift of this size (0.009), however, such a problem is negligible, since the relaxed resolution of the triple axis instrument in the vertical direction corresponds to an integration of around 0.1 reciprocal lattice units along the [1$\overline{1}$0] direction. In addition, the pure sample-rotation method implies that the value of $l$ at the peak is not exactly 2, as anticipated. This is no issue either, since the value of the scattering varies slowly, if at all, with the value of $l$. \cite{Roemer2015}

\begin{table}[h!]
\begin{tabular}{|l|l|l|}
      \hline & \ 2 K, 0 meV, 0 MPa \ & \ 2 K, 0 meV, 30 MPa \ \\ \hline
\ $a$ \ $(10^{-5})$ \      & \ 11.560 $\pm$  1.748  & \ 9.189 $\pm$  1.683     \\
\ $mu$ \     & \ 0.610 $\pm$  0.001    & \ 0.611 $\pm$  0.002     \\
\ $sigma$ \  & \ 0.0095 (fixed)     & \  0.0095 (fixed)    \\
\ $bg$ \ $(10^{-5})$ \     & \ 57.995 $\pm$  0.850  & \ 63.900 $\pm$  0.751 \\ \hline
\end{tabular}
\quad
\begin{tabular}{|l|l|l|}
      \hline & \ 20 K, 0 meV, 0 MPa \ & \ 20 K, 0 meV, 30 MPa \  \\ \hline
\ $a$ \ $(10^{-5})$ \     & \ 10.228 $\pm$  1.735   & \ 4.643 $\pm$  1.812         \\
\ $mu$ \    & \ 0.611 $\pm$  0.002     & \ 0.610 $\pm$  0.006            \\
\ $sigma$ \  & \ 0.0095 (fixed)     & \ 0.0095 (fixed)           \\
\ $bg$ \ $(10^{-5})$ \     & \ 57.435 $\pm$  0.847   & \ 65.366 $\pm$  0.895     \\ \hline
\end{tabular}
\caption{Fitting parameters of the Gaussian fits to the elastic signals shown in Fig.~\ref{fig:2-20elastic}, with $a$ the height of the peak, $mu$ the peak position, $sigma$ the standard deviation (see Table \ref{Tab:fitparameters0MPasummed}) and $bg$ the background. $a$ and $bg$ are given in counts per monitor.}
\label{Tab:fitparameterselastic}
\end{table}

\begin{table}[h!]
\begin{tabular}{|l|l|l|}
      \hline & \ 45 K, 0 meV, 0 MPa \  \\ \hline
\ $a$ \ $(10^{-5})$ \ & \ 2.382 $\pm$  1.659       \\
\ $mu$ \    & \ 0.610 $\pm$  0.006        \\
\ $sigma$ \ & \ 0.0095 (fixed)       \\
\ $bg$ \ $(10^{-5})$ \   & \ 56.068 $\pm$  0.831     \\ \hline
\end{tabular}
\caption{Fitting parameters of the Gaussian fit to the elastic signals shown in Fig.~\ref{fig:45K}, with $a$ the height of the peak, $mu$ the peak position, $sigma$ the standard deviation (see Table \ref{Tab:fitparameters0MPasummed}) and $bg$ the background. $a$ and $bg$ are given in counts per monitor.}
\label{Tab:fitparameterselastic45K}
\end{table}

\begin{table}[h!]
\begin{tabular}{|l|l|l|}
      \hline & \ 20 K, 1 meV, 0 MPa \ & \ 20 K, 1 meV, 30 MPa \ \\ \hline
\ $a$ \ $(10^{-6})$ \ & \ 2.881 $\pm$  0.359    & \ 3.326 $\pm$  0.365     \\
\ $mu$ \    & \ 0.613 $\pm$  0.002     & \ 0.622 $\pm$  0.002      \\
\ $sigma$ \ & \ 0.018 $\pm$  0.003     & \ 0.018 (fixed)       \\
\ $bg$ \ $(10^{-6})$ \   & \ 7.889 $\pm$  0.175   & \ 8.720 $\pm$  0.152   \\ \hline
\end{tabular}
\caption{Fitting parameters of the Gaussian fits to the inelastic signals shown in Fig.~\ref{fig:inelastic}, with $a$ the height of the peak, $mu$ the peak position, $sigma$ the standard deviation (see Table \ref{Tab:fitparameters0MPasummed}) and $bg$ the background. $a$ and $bg$ are given in counts per monitor.}
\label{Tab:fitparametersinelastic}
\end{table}

\pagebreak
\bibliography{biblio_LBCO}
